\documentclass[conference]{IEEEtran}

\usepackage{mathptmx} % This is Times font
\usepackage{amsmath,amssymb,amsfonts}
\usepackage{algorithmic}
\usepackage{graphicx}
\usepackage{textcomp}
\usepackage{fancyhdr}
\usepackage[normalem]{ulem}
\usepackage[hyphens]{url}
\usepackage{cite}
\usepackage[final]{microtype}
\usepackage[keeplastbox]{flushend}
\usepackage{booktabs}
\usepackage{siunitx}

\usepackage{listings}
\usepackage[dvipsnames]{xcolor}
\usepackage{xargs}
\usepackage{xspace}
\usepackage{multirow}
\usepackage{tabularx}
\usepackage{footnote}
\makesavenoteenv{table*}
\makesavenoteenv{tabular}

\lstdefinestyle{customc}{belowcaptionskip=1\baselineskip,
  breaklines=true,
  frame=L,
  xleftmargin=\parindent,
  language=C,
  showstringspaces=false,
  basicstyle=\footnotesize\ttfamily,
  keywordstyle=\bfseries\color{green!40!black},
  commentstyle=\itshape\color{purple!40!black},
  identifierstyle=\color{blue},
  stringstyle=\color{orange},
  numbers=left,
  numberstyle=\tiny,
  numbersep=7pt,
  captionpos=b,
}

\lstdefinestyle{customasm}{belowcaptionskip=1\baselineskip,
  frame=L,
  xleftmargin=\parindent,
  language=[x86masm]Assembler,
  basicstyle=\footnotesize\ttfamily,
  commentstyle=\itshape\color{purple!40!black},
}

\fancypagestyle{firstpage}{
  \fancyhf{}
  
  \fancyfoot[C]{\thepage}
}

\title{Load Driven Branch Predictor (LDBP)}
\begin{document}

\author{\IEEEauthorblockN{Akash Sridhar}
\IEEEauthorblockA{\textit{Computer Science and Engineering} \\
\textit{University of California, Santa Cruz}\\
aksridha@ucsc.edu}
\and
\IEEEauthorblockN{Nursultan Kabylkas}
\IEEEauthorblockA{\textit{Computer Science and Engineering} \\
\textit{University of California, Santa Cruz}\\
nkabylka@ucsc.edu}
\and
\IEEEauthorblockN{Jose Renau}
\IEEEauthorblockA{\textit{Computer Science and Engineering} \\
\textit{University of California, Santa Cruz}\\
renau@ucsc.edu}
}

\maketitle
\thispagestyle{firstpage}
\pagestyle{plain}

%%%%%% -- PAPER CONTENT STARTS-- %%%%%%%%

\begin{abstract}

Branch instructions dependent on hard-to-predict load data are the leading branch
 misprediction contributors. Current state-of-the-art history-based branch predictors have
 poor prediction accuracy for these branches. Prior research backs this observation by
 showing that increasing the size of a 256-KBit history-based branch predictor to its
 1-MBit variant has just a 10\% reduction in branch mispredictions.

We present the novel Load Driven Branch Predictor (LDBP) specifically targeting
 hard-to-predict branches dependent on a load instruction. Though random load data
 determines the outcome for these branches, the load address for most of these data
 has a predictable pattern. This is an observable template in data structures like
 arrays and maps. Our predictor model exploits this behavior to trigger future loads
 associated with branches ahead of time and use its data to predict the branch's
 outcome. The predictable loads are tracked, and the precomputed outcomes of the branch
 instruction are buffered for making predictions. Our experimental results show that
 compared to a standalone 256-Kbit IMLI predictor, when LDBP is augmented with a 150-Kbit
 IMLI, it reduces the average branch mispredictions by 20\% and improves average IPC by
 13.1\% for benchmarks from SPEC CINT2006 and GAP benchmark suite.

\end{abstract}

\section{Introduction}
\label{sec:intro}

% Branch Prediction is important

%Branch mispredictions and data cache misses are the two of the most significant
%factors limiting single-thread performance in modern microprocessors.
%Improving the branch prediction accuracy has a several benefits: Improves IPC
%by reducing the number of flushed instructions; reduces the power dissipated
%through the execution of instructions taking the wrong path of the branch;
%increases the memory level parallelism by allowing to go deeper and triggering
%more memory operations.

Branch mispredictions and data cache misses are the two most significant factors
 limiting single-thread performance in modern microprocessors. Improving the branch
 prediction accuracy has several benefits. First, it improves IPC by reducing the
 number of flushed instructions. Second, it reduces the power dissipation incurred through
 the execution of instructions taking the wrong path of the branch. Third, it
 increases the Memory Level Parallelism(MLP), which facilitates a deeper instruction
 window in the pipeline and supports multiple outstanding memory operations.

% Many branch predictors but TAGE and Perceptron derivates are king

%Current branch prediction championships and CPU designs use either
%perceptron-based
%predictors~\cite{jimenez2001dynamic,jimenez2003fast,ishii2007fused,jimenez2002neural}
%or TAGE-based predictors~\cite{seznec2006case,seznec2011new}. These predictors
%may use global and local history, and a statistical corrector to further
%improve performance.  The TAGE-SC-L~\cite{seznec2016tage} combines several of these
%techniques and was the winner of the last branch prediction championship (CBP-5). The
 %same championship~\cite{cbp5} shows that going from a large 64KBit branch predictor
 %to unlimited size, the MPKI is reduced from 3.986 to 2.596.

Current branch prediction championships and CPU designs use either perceptron-based
predictors~\cite{jimenez2001dynamic}~\cite{jimenez2003fast}~\cite{ishii2007fused}~\cite{jimenez2002neural} or
TAGE-based predictors~\cite{seznec2006case}~\cite{seznec2011new}. These predictors may use global
 and local history, and a statistical corrector to further improve performance.  The
 TAGE-SC-L~\cite{seznec2016tage}, which is a derivative of its previous implementation from
 Championship Branch Prediction (CBP-4)~\cite{seznec2014tage}, combined several of these
 techniques and was the winner of the last branch prediction championship (CBP-5). Numbers
 from CBP-5~\cite{seznec2016tage}~\cite{seznec2016exploring} shows that scaling from a 64-Kbit TAGE
 predictor to unlimited size, only yields branch Mispredictions per Kilo Instructions (MPKI)
 reduction from 3.986 to 2.596.

% Baseline for this work is IMLI 256KBit

%Most of the current processors like AMD Zen2, ARM A72, Intel Skylake use some
%TAGE variation. TAGE-like predictors are excellent, but there are still many
%difficult-to-predict branches. Seznec\cite{seznec2015inner} studied the
%prediction accuracy of a 256-Kbit TAGE predictor and a no storage limit TAGE.
%The 256-Kbit TAGE had only 10\%\textbf{CHECK VALUE} more mispredictions than its infinite size
%counterpart. This is consistent with latest Zen2 CPU with 256Kbit TAGE-like
%predictor~\cite{zen2_hotchips}. For this work, we use the award winning 256Kbit
%IMLI with loop predictor, statistical corrector, and local history as our
%baseline system.

Most of the current processors like AMD Zen 2, ARM A72 and Intel Skylake use some TAGE variation
 branch predictor. TAGE-like predictors are excellent, but there are still many difficult-to-predict
 branches. Seznec~\cite{seznec2014tage}~\cite{seznec2016tage} studied the prediction accuracy of a 256-Kbit
 TAGE predictor and a no storage limit TAGE. The 256-Kbit TAGE had only about 10\% more mispredictions
 than its infinite size counterpart. The numbers mentioned above would reflect the prediction accuracy
 of the latest Zen 2 CPU~\cite{hotchips2019zen2} using a 256-Kbit TAGE-based predictor. For this work, we
 use the 256-Kbit TAGE-GSC + IMLI~\cite{seznec2015inner}, which combines the global history components of
 the TAGE-SC-L with a loop predictor and local history as our baseline system.

% Still opportunities left, predictors fail when dependent on loads that are unpredictable

%Recent work~\cite{lin2019branch} shows that even current state-of-the-art
%branch predictors has significant left opportunities.
%\cite{gao2008address,farooq2013store} have tried to address different types of
%hard-to-predict branches. A key observation of these works is that most
%branches that state-of-the-art predictors fail to capture are branches that
%depend on a recent load.  If the data loaded is challenging to predict,
%TAGE-like predictors have a low prediction accuracy as these patterns are
%arbitrary and too larger to be captured.

Recent work~\cite{lin2019branch} shows that even though the current state-of-the-art branch predictors
have almost perfect prediction accuracy, there is scope for gaining significant performance by fixing
the remaining mispredictions. The core architecture could be tuned to be wider if it had the support of
better branch prediction, which could potentially offer more IPC gains. Prior
works~\cite{gao2008address,farooq2013store} have tried to address different types of
hard-to-predict branches. A vital observation of these works is that most branches that
state-of-the-art predictors fail to capture are branches that depend on a recent load. If
the data loaded is challenging to predict, TAGE-like predictors have a low prediction accuracy
as these patterns are arbitrary and too large to be captured.

%To showcase the extent of speedup exploitable, we compared the IPC
%gains of an AMD Zen 2 core having oracle (prefect) branch prediction compared to a baseline core
%having a 256-Kbit IMLI predictor. Figure~\ref{fig:oracle_speedup} shows that perfect branch prediction
%can achieve an average IPC gain of 82.79\% across the different benchmarks tested. The oracle configuration
%yielded this IPC improvement without any architectural optimizations to the Zen 2 architecture. This
%architecture could be tuned to be wider if it had the support of better branch prediction, which could
%potentially offer more IPC gains.

%\begin{figure}[htb]
%	\centering
%  \includegraphics[width=1\linewidth]{plots/oracle_speedup.png}
%  \caption{\textcolor{red}{Rerun plot} IPC gain for oracle branch predictor over baseline 256-KBit IMLI predictor}
%\label{fig:oracle_speedup}
%\end{figure}

% Predictable load address vs predictable load data

%The key observation/contribution of this paper is that although the load data
%may be difficult to predict, the load address may be easy to predict. If a
%branch uses a difficult to predict load data, the branch is going to be
%difficult to predict. If the load address is predictable, it is possible to
%"prefetch" the load ahead of time, and use the real data value in the branch
%predictor.

The critical observation/contribution of this paper is that although the load data feeding
 a load-dependent branch may be random, the load address may be highly predictable for some
 cases. If the branch operand(s) are dependent on arbitrary load data, the branch is going to
 be difficult to predict. If the load address is predictable, it is possible to "prefetch" the
 load ahead of time, and use the actual data value in the branch predictor.

% LDBP
%Based on the previous observation, we propose to combine the stride prefetcher
%with a new type of the branch predictor to trigger loads ahead of time, and
%feed those loads to the branch predictor. Then, when the branch is "predicted"
%at fetch time, the proposed predictor can have a very high accuracy even with
%random data. The predictor is only active for branches that depend on loads with
%predictable addresses. Otherwise, the default IMLI predictor is used. The proposed
%predictor is called Load Driven Branch Predictor (LDBP).

Based on the previous observation, we propose to combine the stride address
predictor~\cite{fu1992stride} with a new type of the branch predictor to
trigger loads ahead of time and feed the load data to the branch
predictor. Then, when the corresponding branch gets fetched, the proposed
predictor will have a very high accuracy even with random data. The predictor
is only active for branches that have low confidence with the default predictor
and depends on loads with predictable addresses. Otherwise, the default IMLI
predictor performs the prediction. The proposed predictor is called Load
Driven Branch Predictor (LDBP).

% New class of predictors

%LDBP is an implementation of a new class of branch predictors that combine
%load(s) and branches to perform prediction. A load-assisted branch predictor
%allows having near-perfect branch prediction accuracy over random data as long
%as the load address is predictable. It is still a prediction because there are
%possibilities of coherence or other forwarding issues that can make it
%difficult to guarantee the results.

LDBP is an implementation of a new class of branch predictors that combine load(s) and
 branches to perform prediction. This new class of load-assisted branch predictors allows having
 near-perfect branch prediction accuracy over random data as long as the load address
 is predictable. It is still a prediction because there are possibilities of coherence
 or other forwarding issues that can make it difficult to guarantee the results.

% High level overview and naming for paper

%LDBP does not require software changes. It tracks the backward slice from the
%branch to that terminates in a set of loads. If all the loads have a
%predictable address, and the slice is small enough to be computed, the slice is
%remembered. Next time that the branch retires, it will start to trigger
%prefetches ahead of time so that the next time that it is fetched the
%precomputed slice result could be used.  Through the rest of this paper, we
%will refer to the predictable address load that has a dependency with a branch
%as a trigger load and its dependent branch as load-dependent branch.

LDBP does not require software changes or modifications to the ISA. It tracks the
 backward code slice starting from the branch and terminating at a set of one or more
 loads. If all the loads have a predictable address, and the slice is small enough to
 be computed, LDBP keeps track of the slice. When the same branch retires again, it will
 start to trigger future loads ahead of time. The next fetch of this branch uses the precomputed
 slice result to predict the branch outcome. Through the rest of this paper, we will refer
 to the load(with predictable address) that has a dependency with a branch as a trigger load
 and its dependent branch as a load-dependent branch.

% Example traverse table

%To show a code example that can benefit from LDBP, let us consider code
% Listing~\ref{lst:asm}.  This is a loop that iterates over a vector that counts
% the number of negative values.  If the data is random, the even the best branch predictors
%are not going to capture the pattern. LDBP has near perfect branch prediction
%because the trigger load has predictable address. LDBP triggers loads ahead of
%time, computes the branch backward slice, and stores the results. When the
%branches are fetched, the precomputed result is used. The result is that for a
%Zen 2 like core with a 256-Kbit IMLI predictor, the IPC goes from \textbf{X} to
%\textbf{Y}.

\begin{minipage}{0.45\textwidth}
\centering
\begin{lstlisting}[frame=single, caption={Vector traversal code snippet example}, label=lst:asm]
addi	a5,a5,4 //increments array index
  .
  .
lw	a4,0(a5)//loads data from array
bnez	a4,1043e <main+0x44>
\end{lstlisting}
\end{minipage}

We will explain a simple code example that massively benefits from LDBP. Let us consider
 a simple kernel that iterates over a vector having random 0s and 1s to find values greater
 than zero. The branch with most mispredictions in this kernel has the assembly sequence shown in
 Listing~\ref{lst:asm}. As we are traversing over a vector, the load addresses here are
 predictable, even though the data is completely random. TAGE fails to build these branch
 history patterns due to the dependence of the branch outcome on irregular data patterns. LDBP
 has near-perfect branch prediction because the trigger load (line 4) has a predictable
 address. LDBP triggers loads ahead of time, computes the branch-load backward slice, and
 stores the results. The branch uses the precomputed outcome at fetch. When we augment LDBP
 to a Zen 2 like core with a 256-Kbit IMLI predictor, the IPC improves by ~2.6x times.

% Key challenge

%LDBP key challenge is send trigger loads to reach the branch predictor before
%the corresponding load-dependent branch is fetched. This is achieved by
%leveraging the stride prefetcher. When a trigger load has a predictable load
%address, the stride prefetcher learns the delta with a high confidence. When
%the trigger load retires, it triggers prefetches with enough distance to cover
%the in-flight instructions and the memory latency. As Section~\ref{sec:model}
%shows, this can be achieved with very small structures and a small overhead.

In general, a load-dependent branch immediately follows a trigger load in program order. Due
 to the narrow interval between these two instructions, the load data will not be available when
 the branch is fetched. Therefore, if this load yields a stream of random data across iterations, LDBP
 will have a very slim chance of making a correct prediction. To address this issue, we ensure the
 timeliness of the trigger loads in our setup. The key challenge is to make sure that the trigger
 load execution is complete before the corresponding load-dependent branch reaches fetch. By leveraging
 the stride predictor, we can ensure trigger load timeliness. When a branch retires, a read request
 for a trigger load is generated. Owing to the high predictability of their address, trigger load
 requests future addresses in advance. These requests have sufficient prefetch distance to cover the
 in-flight instructions and variable memory latency. As Section~\ref{sec:ldbp} shows, this
 can be achieved with very small structures incurring little hardware overhead.

% Benchmarks selected

%To evaluate the results, we use GAPB~\cite{beamer2015gap} and the SPEC2006
%benchmarks that less than 95\% misprediction accuracy in our IMLI baseline.
%GAPB is a collection of graph algorithm benchmarks. This is one of the highest
%performance benchmarks available, and graphs are know to be severely limited by
%the branch prediction accuracy. Improvements in their performance have a
%significant impact in many areas like search and AI.

To evaluate the results, we use the GAP benchmark suite~\cite{beamer2015gap} and the SPEC2006 integer
benchmarks~\cite{henning2006spec} having less than 95\% prediction accuracy on our IMLI baseline. GAP
is a collection of graph algorithm benchmarks. This is one of the highest performance benchmarks
available, and graphs are known to be severely limited by branch prediction accuracy. We integrated an
81-Kbit LDBP to the baseline 256-Kbit IMLI predictor. Results show that LDBP fixes the topmost
mispredicting branches for more than half of the benchmarks analyzed in this study. Compared to the
baseline predictor, LDBP with IMLI decreases the branch MPKI by 22.7\% on average across all
benchmarks. Similarly, the combined predictor has an average IPC improvement of 13.7\%. LDBP also
eases the burden on the hardware budget of the primary predictor. When combined with a 150-Kbit IMLI
predictor, the branch mispredictions come down by 20\%, and the performance gain scales by 13.1\%
compared to the 256-Kbit IMLI, for a 9.7\% lesser hardware allocation.

% Preliminary results
% The preliminary results show \textbf{\textcolor{red}{XXX}}.

% Rest of the paper
The rest of the paper is organized as follows: Section~\ref{sec:ldbp} describes the LDBP mechanism
and architecture. Section~\ref{sec:setup} reports our evaluation setup methodology. Benchmark
analysis, architecture analysis, and results are highlighted in
Section~\ref{sec:evaluation}. Section~\ref{sec:related} presents related works. Section~\ref{sec:conclusions}
concludes the paper.

\section{Load Driven Branch Predictor}
\label{sec:ldbp}

\subsection{Load-Branch Chains}

The core principle of LDBP involves the exploitation of the dependency between
load(s) and a branch in a load-branch chain. In this sub-section, we will explain
load-branch chains in detail. LDBP needs to capture the backward
slice\cite{moshovos.ics01} of operation sequence starting from the branch. The exit
point of this slice must be a load with a predictable address or a trivially computable
operation like a load immediate operation.

\begin{figure}[htb]
	\centering
    \includegraphics[width=0.9\linewidth]{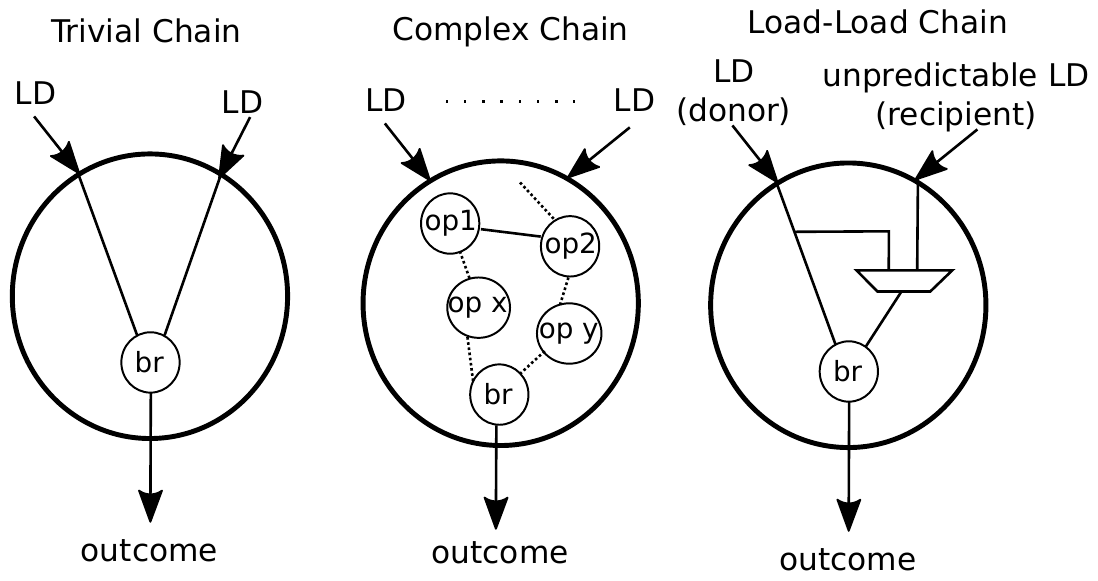}
    \caption{Generic load-branch chain starts with predictable loads and terminates with a branch.}
\label{fig:chain}
\end{figure}

As shown in Figure~\ref{fig:chain}, we classify load-branch chains into 3 different
types: trivial, complex and load-load chain. In a trivial chain, the branch has a single source
operand (like a $bnez$ instruction) or two source operands, and it has a direct dependency with a predictable
load. No intermediate instructions modify the load data in this chain.

In a complex chain, all the branch inputs terminate with a predictable load or a load
immediate. A complex chain includes at least one predictable load, one or more simple
arithmetic operations, and it concludes with the branch. The LDBP framework does not
track complex ALU operations, and any chain with such an operation is invalidated. We
will explain load-load chain in Section~\ref{sec:evaluation}.

A load-branch chain has two main constraints: (1) the maximum number of operations between the load
and the branch, (2) the maximum number of input loads. For example, a chain can have five simple ALU
operations before the branch. It means that a Finite State Machine (FSM) of the chain needs six cycles
to compute the branch result. From the benchmarks we analyzed, we found that a considerable proportion of
hard-to-predict branches are part of a trivial load-branch chain.

\subsection{LDBP Architecture}

In this sub-section, we explain the LDBP architecture. As LDBP works in conjunction with the
primary branch predictor, its architecture aims at being simple, timely, spectre-safe, and
having low power overhead. The LDBP architecture is dissected into two sub-blocks: one block
attached to the core’s retirement stage and another block at the fetch stage. From an abstract
level, the retirement block detects potential load-branch chains, creates backward slices from
the branch to its dependent load(s), and generates trigger loads. On the other hand, the fetch
block uses the backward slices to build FSMs of the program sequence and computes the outcome
of load-dependent branches using the executed trigger load data.

%From a high level, we can break the LDBP flow into three steps. One, identifying
%potential load-branch chains and constructing its associated backward slices at
%retirement. Two, generate trigger loads to be used at the front end. Three, build
%FSMs using the backward slice and use the trigger load data in these FSMs to predict
%the outcome of the load-dependent branch at the fetch stage.

%Several alternative designs can have different trade-offs in leveraging the 
%load-branch chains. This section explains a potential architecture that aims at
%being simple, low power overhead, and timely in generating the triggered loads. 

%The proposed LDBP architecture has a block attached to the core retirement stage
%and another attached to the fetch stage. From a high level, the retirement block
%detects potential load-branch chains, creates backward slices from difficult to
%predict branches, and sends trigger loads to execute. The fetch block computes 
%the branch outcomes when all the needed triggered loads reach the fetch. It does
%so by using the backward slices and updating the potential future trigger branches.

%To explain the LDBP, we explain the LDBP retirement and fetch block fields, and 
%then we proceed to explain how they are updated.

\subsubsection{LDBP Retirement Block}

\begin{figure}[htb]
	\centering
  \includegraphics[width=0.95\linewidth]{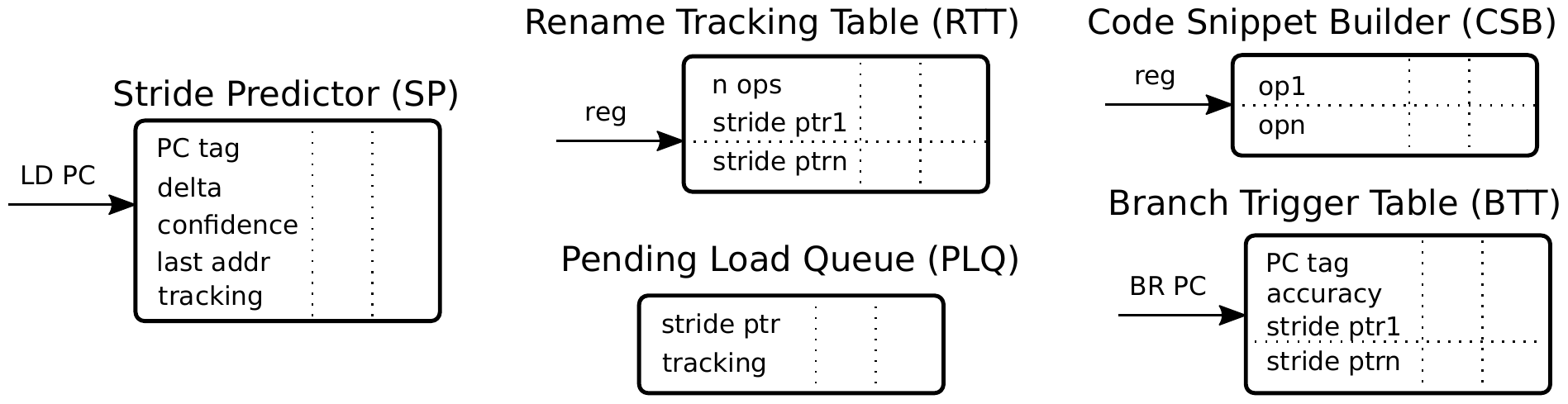}
    \caption{LDBP Retirement Block - Fields in each index of the tables are marked in the figure.}
\label{fig:retire_ldbp}
\end{figure}

A naive LDBP retirement block could consume significant power detecting and building
backward slices all the time. To avoid this substantial power overhead, we leverage the
stride address predictor that exists in many modern microarchitectures to detect predictable
loads. In addition to that, LDBP attempts to identify a load-branch chain only when the
load is predictable, and the associated branch has low confidence with the default
predictor (in our case, it is IMLI predictor). Figure~\ref{fig:retire_ldbp} shows the
tables/structures associated with the retirement block.

%A naive LDBP retirement block could consume significant power detecting and building
%backward slices all the time. To avoid this power overhead, we leverage the stride
%prefetcher that exists in most cores to detect predictable loads, and we only
%initiate to build a potential load-branch chain when the branch predictor (TAGE) 
%has a low confidence and it has a branch miss prediction.

{\bf \noindent Stride Predictor (SP)}:
The retiring load PC indexes the Stride Predictor table. This table has five fields. They
are the PC tag (\emph{sp.pctag}), the address of the last retired load
(\emph{sp.lastaddr})~\footnote{Stride predictor can store partial load addresses to save
space}, the load address delta (\emph{sp.delta}), a delta confidence counter
(\emph{sp.confidence}) and a tracking bit to indicate if a given load PC is tracked as a
part of a load-branch chain (\emph{sp.tracking}).

The updating policy of the confidence counter varies across different stride predictors. Standard
practice involves increasing the counter each time the delta repeats and decreasing it each time
the delta changes. This approach may skew the confidence either way. Ideally, increasing the counter
by one and reducing it by a higher value minimizes the bias. A tracked load (with the
\emph{sp.tracking} set) can trigger only when its confidence counter is saturated.

{\bf \noindent Rename Tracking Table (RTT)}:
The Rename Tracking Table detects and builds dependencies in the load-branch chains. The
retiring instruction's logical register indexes the RTT\@. Each table entry has a saturating
counter to track the number of operations (\emph{rtt.nops}) in a load-branch chain and a
pointer list to track Stride Predictor entries (\emph{rtt.strideptr}). The number of entries
in the pointer list depends on the number of loads supported by LDBP\@. If a chain consists
of 2 loads and 4 arithmetic operations before the branch, we need 3 bits to track these
six operations and two entries on the pointer list.

%{\bf \noindent Rename Tracking Table (RTT)}: The Rename Tracking Table (RTT) 
%builds and detects load-branch chains.  The RTT is indexed by the retiring 
%instruction's logical register. Each entry has a saturating number of operations 
%(\emph{rtt.nops}) and several pointers to the stride prefetcher entry 
%(\emph{rtt.strideptr}).

%The number of pointers to the stride prefetcher depends on how many loads are 
%supported. If a slice supports 2 trigger loads and 4 operations per load-branch
%chain, we need 3 bits for the number of operations, and 2 entries for stride
%pointers. If the stride prefetcher tracks 256 PCs (8 bit index, and 1 bit valid),
%we need 21 bits total per entry.

{\bf \noindent Branch Trigger Table (BTT)}:
The Branch Trigger Table links a branch with its associated loads and intermediate
operations. The retiring branch PC indexes the BTT. Each entry has the following
fields: the branch PC tag (\emph{btt.pctag}), the list of associated loads (copied
from the Stride Predictor pointer list from the RTT table (\emph{btt.strideptr})), and
a 3-bit accuracy counter to track  LDBP's accuracy for this branch
(\emph{btt.accuracy}). If the accuracy counter reaches zero, the BTT entry gets
cleared, and \emph{sp.tracking} bits of the loads in \emph{btt.strideptr} are reset. A
BTT entry is allocated only when a load-branch chain satisfies the following three
conditions: (1) the loads in the chain are predictable; (2) the retiring
branch has low confidence with IMLI; (3) number of loads and number of operations in
the chain is within the permissible threshold.

%{\bf \noindent Branch Trigger Table (BTT)}: The BTT is indexed by the retiring 
%branch PC. The fields have the branch PC tag (\emph{btt.pctag}), all the stride
%prefetcher pointer from the RTT table (\emph{btt.strideptr}), and the FSM pointer
%to execute the branch backward slice (\emph{btt.fsmptr}).

{\bf \noindent Code Snippet Builder (CSB)}:
The CSB tracks the operation sequence of a load-branch chain for each logical register. Each
entry on this table is a list of operations (\emph{csb.ops}). The CSB entry is updated only
when a new BTT entry gets allocated. This prerequisite ensures that the CSB is not polluted
and minimizes power overhead. There are several works in the academic literature about building
backward slices~\cite{moshovos.ics01}. We use a table indexed by the retiring logical register
(similar in behavior to an RTT). It copies the chain of operations starting from the load and
terminating with the branch. Initially, we considered the possibility of combining the CSB with
the RTT but dropped the idea considering the additional power dissipation this would incur. The
CSB entries are only needed when a new BTT entry is populated(when a load-branch chain is
established), and it would not make sense to integrate it with the RTT\@.

%{\bf \noindent Code Snippet Builder (CSB)}: It has the load-branch chain for each
%logical register. Each entry is a list of operations (\emph{csb.ops}). The CSB is
%activated only when a new entry is added to the BTT. There are several works at the
%literature to build backward slices~\cite{slice}. A simple solution is to have a
%table indexed by the retiring logical register address that behaves like the RTT
%but copies the chain of operations and the Stride Prefetcher entry. It is possible
%to combine the RTT with the CSB, but this will have a higher power overhead due to
%the extra CSB entries that are only needed when a new BTT entry is populated.

\begin{figure}[htb]
	\centering
    \includegraphics[width=0.95\linewidth]{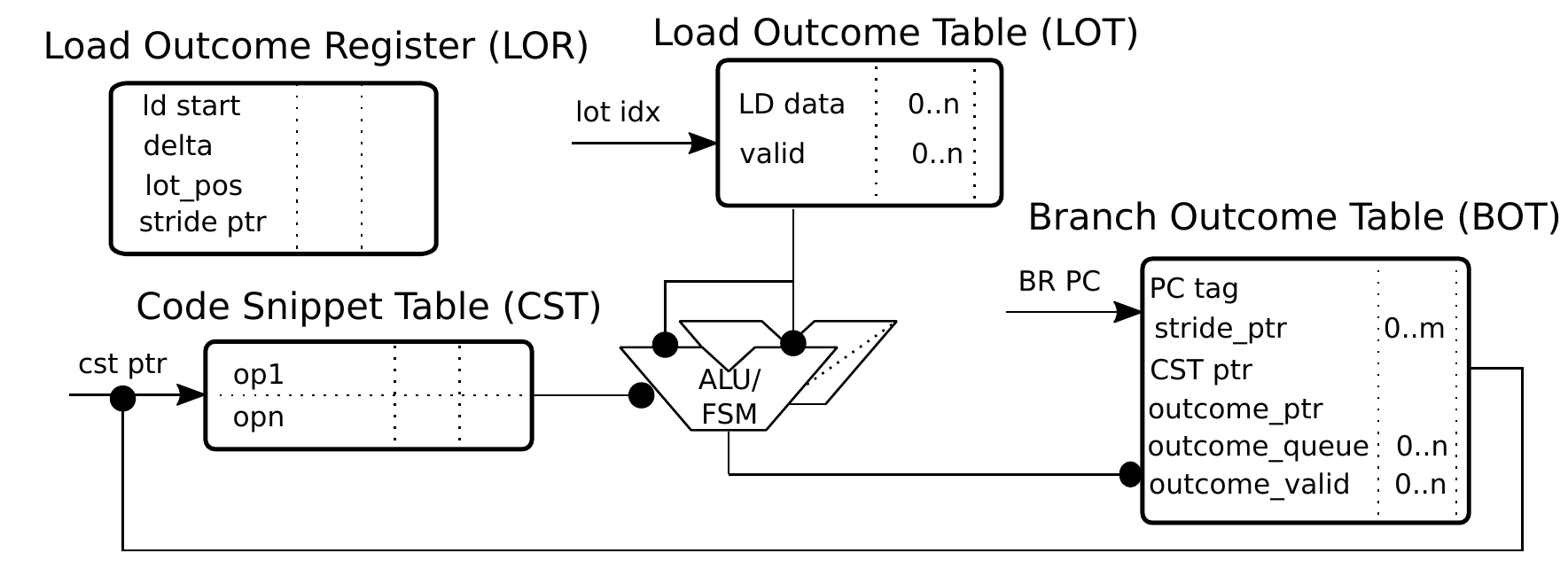}
    \caption{LDBP Fetch Block - Fields in each index of the tables are marked in the figure.}
\label{fig:fetch_ldbp}
\end{figure}

\begin{figure*}[ht]
	\centering
  \includegraphics[width=0.75\textwidth]{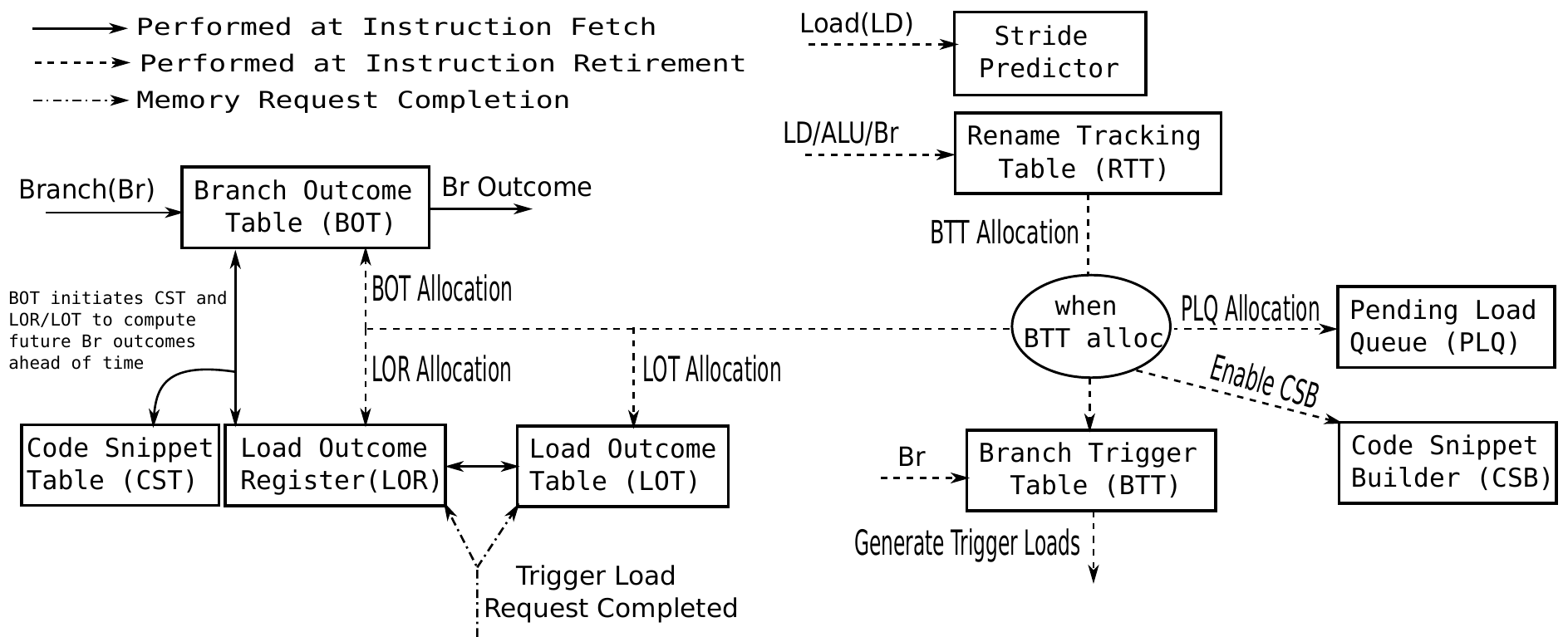}
  \caption{LDBP Flow - Interaction between Fetch and Retire Block.}
\label{fig:ldbp_flow}
\end{figure*}

{\bf \noindent Pending Load Queue (PLQ)}:
The tables/structures mentioned above is sufficient to detect and build load branch chains. The
PLQ acts as a buffer and stores the Stride Predictor pointer list (\emph{plq.strideptr}) associated
with a load-branch chain. It tracks whether the last retired load had a change in delta
(\emph{plq.tracking}). If there is a change, it notifies the retire block to stop triggering potentially
incorrect loads. Generally, loads generate prefetches when it retires. But, in our setup, we delay the
trigger load generation until the branch retires to ensure correctness in trigger load generation. The
PLQ ensures that the BTT gets notified about any change in the retiring load's delta before it triggers
any loads. As shown in Figure~\ref{fig:ldbp_flow}, PLQ allocates entries during BTT allocation.

%{\bf \noindent Pending Load Queue (PLQ)}: The previous structures are enough to
%detect and build load-branch chains. As each load retires, we should send the
%load to execute using the retiring load address plus the delta and a distance.
%Instead, of tracking distance, LDBP delays the load execution until the trigger
%branch is executed. The PLQ keeps the stride pointer (\emph{plq.striptr})
%and whether the last retired load had a change of delta (\emph{plq.tracking}).

\subsubsection{LDBP Fetch Block}

The LDBP fetch block is responsible for accumulating trigger load results and computing the
branch outcomes. Figure~\ref{fig:fetch_ldbp} shows the tables used by the fetch block, the
registers associated with tracking loads, and the ALU used to compute the branch outcome for
load-branch chains.

%The fetch block is responsible for accumulating load results and computing the
%branch outcomes. To speedup the branch resolution, the computations are done ahead
%of the trigger branch fetch.  Figure~\ref{fig:fetch_ldbp} shows the 3 tables used
%by the LDBP fetch block, the registers associated to track loads, and the ALU used
%to compute the branch outcome for simple chains.

{\bf \noindent Load Outcome Table (LOT) and Load Outcome Registers (LOR)}:
The combination of LOR and LOT stores trigger load data, which could be consumed by future
branches. The \emph{lor.ldstart} is the starting load address of the range, and it is updated
after every branch fetch. The \emph{lor.delta} field tracks the load address delta of each load
tracked by LOR (\emph{lor.strideptr}). The \emph{lor.lot\_pos} field marks the data to be used by the current
branch, and it helps to queue incoming data in an appropriate LOT index. The \emph{lot.valid} bit
gets set when the trigger load associated with that entry finishes execution.

%The combination of LOR and LOT stores trigger load data, which could be consumed by future
%branches. The LOR keeps track of a range of load addresses whose data could be potentially
%useful for the current and future branches. The LOT caches the data associated with the
%addresses tracked by LOR. Any trigger load address outside the address range is deemed
%useless, and the LOT does not cache its data. Each LOR entry has an associated LOT entry. Each
%LOT entry has an $n$-entry load data queue (\emph{lot.ld\_data}) and valid bit queue((\emph{lot.valid})).

The LOR keeps track of a range of load addresses whose data could be potentially useful for the
current and future branches. The LOT caches the data associated with the addresses tracked by
LOR. Each LOR entry has an associated LOT entry. Each LOT entry has an n-entry load data queue
(\emph{lot.ld\_data}) and valid bit queue (\emph{lot.valid}). The ending address tracked by LOR
is $lor.ldstart + n * lor.delta$. Any trigger load address outside the address range is deemed
useless, and the LOT does not cache its data.

%The \emph{lor.ldstart} is the starting load address of the range, and it is updated after
%every branch fetch. The \emph{lor.delta} field tracks the address delta of the load
%tracked by LOR (\emph{lor.strideptr}). \emph{lor.lotpos} marks the data to be consumed by the
%current branch, and it helps to queue incoming data in an appropriate LOT index. The
%\emph{lot.valid} bit gets set when the trigger load associated with that entry finishes execution.

%{\bf \noindent Load Outcome Table (LOT) and Load Outcome Registers (LOR)}: For
%many benchmarks, monitoring 2 loads is important. The fetch block has to remember
%the values of loads before computing the trigger branch outcome. The combination
%of LOT and LOR remember the load outcomes from triggered loads that may be used
%in future fetch. There is a maximum number of trigger loads that can be monitored
%at the same time. For each trigger load, the need a LOR pair of registers to
%remember the next load start address (\emph{lor.ldstart}) and the delta
%(\emph{lor.delta}) detected by the retirement block. The value of the triggered
%load is stored at the LOT.

{\bf \noindent Branch Outcome Table (BOT):}
The branch PC indexes the BOT at fetch (\emph{bot.pctag}). As shown in Figure~\ref{fig:ldbp_flow}, the
BOT has two main tasks. One, use the pre-computed branch outcome to predict at the fetch stage. Two, initiate
the Code Snippet Table to compute the outcome for future branches.

Each BOT entry has a queue of 1-bit entries holding the branch outcome (\emph{bot.outcome\_queue}). The
length of this queue is equivalent to the number of entries in the \emph{lot.ld\_data} queue. The
\emph{bot.outcome\_ptr} points to the current BOT outcome queue entry to be used by the incoming
branch instruction. BOT uses the outcome if the corresponding \emph{bot.valid} bit is set. The
\emph{bot.strideptr} has the list of loads associated with the branch. The Code Snippet FSM uses
this field to pick appropriate load(s) from the LOR/LOT and the CST pointer (\emph{bot.cstptr}) to
compute the branch outcome.

{\bf \noindent Code Snippet Table (CST):}
The Code Snippet Table (CST) is responsible for executing the branch backward slice to compute the branch
outcome. A CST entry is allocated during BOT allocation. The CST feeds the FSMs with the operation sequence
of the load-branch chain. When all the trigger load data associated with the trigger branch are available, the
FSM executes the code snippet to completion at the rate of one ALU operation per cycle. When large backward
slices are supported, more FSMs are needed to reduce contention. The contention happens when all the FSM are
busy. In this case, the branch outcome gets delayed until an FSM  is free. As the BOT only tracks a small number
of trigger branches, a similar-sized CST is sufficient.

\subsection{LDBP Flow}

Figure~\ref{fig:ldbp_flow} shows the interaction between different LDBP components at instruction
fetch and retirement stage. Through the rest of this sub-section, we will look in detail about LDBP behavior.
%The LDBP retire block builds load-branch chain and triggers loads. The LDBP fetch block is in charge of storing provisional trigger load results and computing the branch outcomes ahead of time. Through the rest of this sub-section, we will look in detail about LDBP behavior.

%{\bf \noindent At Load Retirement}:
\subsubsection{Load Retirement}

When a load retires, it updates the Stride Predictor. The \emph{sp.confidence} field is updated depending
upon the load address behavior. The \emph{sp.tracking} for a load gets set at BOT allocation. A BOT entry
allocation implies that a valid load-branch chain is present, and it is necessary to track the loads
in this chain to ensure LDBP correctness.

If the \emph{sp.tracking} is set, the corresponding Stride Predictor index is appended to the Pending
Load Queue table (\emph{plq.strideptr}). The \emph{plq.tracking} bit remains set until there is a change in
delta for the load it tracks.

The retiring load also resets the RTT entry indexed by its destination register. If the \emph{sp.confidence} is
high, the \emph{rtt.nops} is initialized to zero, and the load's pointer from the Stride Predictor is appended
to the \emph{rtt.strideptr}. In case the \emph{sp.confidence} is low, the \emph{rtt.nops} is saturated, and the RTT stride
pointer list is cleared.

%{\bf \noindent At ALU Retirement:}
\subsubsection{ALU Retirement}
A retiring simple ALU operation (like addition) updates the RTT entries pointed by its destination
register. The RTT retrieves \emph{rtt.nops} and \emph{rtt.strideptr} values pointed by its source
registers~\footnote{At most two sources in RISC-V} and accumulates it into the fields indexed by
the destination register. The cumulative \emph{rtt.nops} is represented by Equation
~\ref{eq:rtt_nops}. It is realistically infeasible to track an infinitely large load-branch
chain. So, there is a threshold on the number of operations and the number of loads
supported by LDBP. If these values in the RTT field exceed the limit, the corresponding RTT entry
gets invalidated. For simplicity, we add the number of operands per source
ignoring any potential redundancy in operations.

\begin{equation}\label{eq:rtt_nops}
   rtt[dst].nops = rtt[src1].nops + rtt[src2].nops + 1
\end{equation}

LDBP does not support complex operations like multiplication or floating-point operations. As a
result, when one of these instructions retire, the RTT entry indexed by it is invalidated to ensure
a load-branch chain does not get polluted by complex operations.

\subsubsection{Branch Retirement}
%{\bf \noindent A branch retires with misprediction and misses BTT:}

At cold start, when a branch retires, it indexes the RTT only when it has low confidence with the default
IMLI predictor~\footnote{IMLI is confident when the longest table hit counter is saturated.}. RTT ensures
the validity of the load-branch chain by checking the load count and operation count in the entry indexed
by the branch source(s). BTT entry gets allocated only when all the loads in this chain are predictable.

{\bf \noindent BTT Allocation:}
On BTT allocation, the contents of the \emph{rtt.strideptr} are copied to the BTT stride pointer list. The BTT
accuracy counter (\emph{btt.accuracy}) is initialized to half of its saturation value. The
\emph{sp.tracking} bit for the associated loads are set, and the CSB starts building the code
snippet for this load-branch chain. As shown in Figure~\ref{fig:ldbp_flow}, the BTT allocation
creates a chain reaction by initiating the PLQ allocation, LOR/LOT allocation, and BOT allocation.

Each load associated with the branch has a unique entry during LOR/LOT allocation. Load-associated
metadata from the Stride Predictor populates the LOR fields. The \emph{lor.lot\_pos} is
cleared. Similarly, BOT entry gets reset on allocation, and \emph{btt.strideptr} updates the stride
pointer list on the BOT\@. The branch's PC tag is assigned to \emph{bot.pctag}.

{\bf \noindent BTT Hit:}
On BTT hit, the \emph{btt.accuracy} counter gets incremented if LDBP made a correct prediction, and
the default IMLI predictor mispredicts and vice versa. If this counter reaches zero, the BTT deallocates
the entry and the \emph{sp.tracking} associated with \emph{btt.strideptr} are cleared.

The CSB starts to build the code snippet for the load-branch chain on BOT allocation. After CSB
completes the snippet, on a BTT hit, the code snippet is copied to the CST. The CSB is disabled
after this process.

When the retiring branch hits on the BTT, it reads the corresponding PLQ entries to ensure if the
tracking bit is high for the loads in the \emph{btt.strideptr}. The BTT can trigger load(s) if the
PLQ and LOR track all the associated loads. Equation~\ref{eq:trigger_load} represents the address
of the load triggered. The \emph{lor.ldstart} is incremented by load address delta to ensure better
coverage after every trigger load generation. The \emph{lor.lot\_pos} is incremented when a new load
is triggered. The trigger load distance (\emph{tl\_dist}) and the number of triggers generated for each
load can be tuned to facilitate better load timeliness.

\begin{equation} \label{eq:trigger_load}
  tl\_addr = lor.ldstart + lor.delta * tl\_dist
\end{equation}

There can be scenarios where the load-branch chain might change. It could happen when a different
operation sequence is taken to reach the branch. There are situations where the delta associated
with any of the branch's dependent loads might change, potentially resulting in triggering incorrect
loads. During such occurrences, LDBP flushes the branch entries on the BTT, BOT (and its associated CST
entry), and its corresponding load entries on the LOR/LOT. The tracking bit on the Stride Predictor
and PLQ are reset for these loads. Such an aggressive recovery scheme guarantees higher LDBP accuracy
and reduced memory congestion due to unwanted trigger loads.

%{\bf \noindent A branch retires without misprediction and misses BTT:}
%In this case, we do not do anything.

%{\bf \noindent A branch retires with misprediction and hits BTT:}
%If there is a misprediction, we can sync the retirement and the fetch block because
%no in-flight instruction exists. If the delta has no changed, the data at the 
%LOR/LOT is still valid. The index and load address is adjusted. If the delta has
%changed, the LOR/LOT table is reset with the new target.

{\bf \noindent Trigger Load Completion:}
When a trigger load completes execution, it checks for matching entries on the LOR. There could be
zero or more entries on the LOR, which could have the address range of this completed request. The
address is a match on the LOR entry if it is within the LOR entry's address range and is a factor
of the \emph{lor.delta}. On a hit, the corresponding LOT entry stores the trigger load data in the
\emph{lot.ld\_data} queue, and its valid bit is set. The LOT data queue index is computed using
Equation~\ref{eq:lot_indexa} and~\ref{eq:lot_indexb}.

\begin{subequations}
  \label{eq:lot_index}
  \begin{align}
    lot\_id &= \frac{(tl.addr - lor.ldstart)}{lor.delta} \label{eq:lot_indexa}\\
    lot\_index &= (lor.lot\_pos + lot\_id ) \% lot.ld\_data.size( ) \label{eq:lot_indexb}
  \end{align}
\end{subequations}

%When a trigger load reaches the LDBP
%Fetch Block, it checks the for a matching LOR. There could be zero or more matches.
%For each match, the value is stored at the corresponding LOT table, and the valid
%is set. The LOT table index is computed using the load start address, the delta, 
%and the index with $\frac{\texttt{exe.ldaddress} - \texttt{lor.ldstart}}{\texttt{lor.delta}} - \texttt{lor.index}$.

\subsubsection{Branch at Fetch}
When a branch hits on the BOT at instruction fetch, the \emph{bot.outcome\_ptr} is increased by
one. This is the only value speculatively updated in the LDBP fetch block. When there is a table
flush due to misprediction, load-branch chain change or load delta variation, the \emph{bot.outcome\_ptr}
gets flushed to zero. The BOT outcome queue entry pointed by the \emph{bot.outcome\_ptr} yields
the branch's prediction.

The \emph{bot.cst\_ptr} proactively instigates the computation of future branch outcomes at fetch. The
CST FSMs use the load data values from valid entries on the LOT. Once the outcome is computed, the
corresponding \emph{bot.outcome\_queue} entry gets updated.

%When the load is fetched, the \emph{bot.fsmptr} is used to find the next next set
%of LOT entries that have all the data present. The Code Snippet FSM is triggered
%for the load data.

\subsection{Spectre-safe LDBP}

The LDBP has been designed to avoid speculative updates. The reason is not to
create another source of Spectre-like~\cite{kocher2018spectre} attacks. The LDBP
retirement block is only updated when the instructions are not speculative.
This means that it never has any speculative information and potential
speculative side-channel leak.

The LDBP fetch block is populated only with information from the retirement
block. Even the trigger loads are sent when a safe target branch retires. The
only speculatively updated field is the LOR table, but this table is flushed
after each miss prediction, and the state is rebuilt from the LDBP retirement
block.

In a way, the LDBP is not a new source of speculative leaks because it is only
updated with safe information, and the fields updated speculatively are always
flushed on any pipeline flush. The flush is necessary for performance, not only
for Spectre.  The reason is that when the "number of in-flight" trigger loads
change due to flushes, the LOR must be updated. LDBP structures are not source
of speculative leaks, but the loads in the speculative path can still leak
unless speculative loads are protected like in~\cite{sakalis2019efficient}. The
result is that LDBP is not a new source of speculative leaks like most branch
predictors that gets speculatively updated and not fixed on pipeline flushes.

\subsection{Multiple Paths Per Branch}

The LDBP load-branch slices are generated at run-time, and they can cross branches. As a result, the
same branch can have multiple chains or backward slices. These cases are sporadic in benchmarks from
GAP as they have a large and somewhat regular pattern. Multi-path branches are slightly more common
in the SPEC CINT2006 benchmarks.

The analysis performed as a part of this work shows that branches with multiple slices are not
frequent, and when they happen, they tend to depend on unpredictable loads. Therefore, it is
not a significant cause of concern for LDBP in these cases. Nevertheless, it can be an issue in
other workloads. We leave it as a part of future work and possibly find benchmarks that exhibit
such behavior more predominantly.

%A potential solution to overcome this is to extend the BTT table to behave like a TAGE predictor. By
%this, we mean that the current BTT table looks like a bimodal table indexed by the current branch
%PC. It is possible to have multiple BTT tables, each indexed with a longer history. The longer history
%tables are indexed when the BTT allocation happens for a new slice. Although not evaluated, it makes
%sense to address this issue because the different slices are dependents on past occurrences of the
%branch. We leave it as a part of future work and possibly find benchmarks that exhibit such behavior
%more predominantly.

\section{Simulation Setup}
\label{sec:setup}

We evaluated LDBP using a subset of SPEC 2006 and the GAP Benchmark Suite~\cite{beamer2015gap}. For
SPEC CINT2006, we ran all the benchmarks skipping 8 billion and modeling for 2 billion instructions. Any
benchmark with branch prediction accuracy less than 95\% is used for our evaluation
(\emph{hmmer}, \emph{astar}, \emph{gobmk}). The other benchmarks in the SPEC CINT2006 suite already have
very low MPKI. Therefore, they would not be a true reflection of the impact of LDBP. We run all the GAP
applications with ``\emph{-g 19 -n 30}'' command line input set and instrument the benchmarks to skip
the initialization, as suggested by the developers of GAP. All the benchmarks are compiled with \emph{gcc 9.2}
with \emph{-Ofast -flto} optimization for a RISC-V RV64 ISA.

 \begin{table}[htb]
   \centering
   \small
   %\resizebox{\columnwidth}{!}{%
   \begin{tabular} { l|r|r }
    \textbf{Benchmark} & \textbf{Branch MPKI} & \textbf{IPC} \\
    \toprule
    spec06\_hmmer  & 12.9  & 2.42 \\ 
    spec06\_astar & 14.9 & 0.89  \\
    spec06\_gobmk & 13.1 & 1.49 \\ 
     \midrule
    gap\_bfs & 23.9 & 0.66 \\
    gap\_pr & 4.6 & 1.64 \\
    gap\_tc & 44.5 & 1.07 \\
    gap\_cc & 32.7 & 0.51 \\
    gap\_bc & 22.0 & 1.14 \\
    gap\_sssp & 6.2 & 0.89 \\
    %\hline
    %\textbf{Total LDBP Size} &  & \textbf{82.56} \\
   \end{tabular}
   %}
   \vspace{0.1in}
   \caption[]{Benchmarks used and their MPKI and IPC running baseline 256-Kbit IMLI.}
   \label{tab:setup_table}
 \end{table}

We use ESESC~\cite{esesc} as the timing simulator. The processor configuration is set to closely model an
AMD Zen 2 core~\cite{hotchips2019zen2}. Table~\ref{tab:setup_table} shows the Instructions Per Cycle (IPC) and MPKI for the benchmarks
investigated when running the baseline 256-Kbit IMLI predictor. To match the Zen 2 architecture, the baseline
branch prediction unit has a fast (1 cycle) branch predictor and a slower but more accurate (2 cycle) IMLI
branch predictor. We evaluate the baseline configuration against 1-Mbit IMLI, and different IMLI
configurations (150-Kbit, 256-Kbit, and 1-Mbit) augmented with an 81-Kbit LDBP.

%\todo{add table: 3 cols name, IPC, MPKI}

\begin{figure}[htpb]
\centering
\includegraphics[width=1\linewidth]{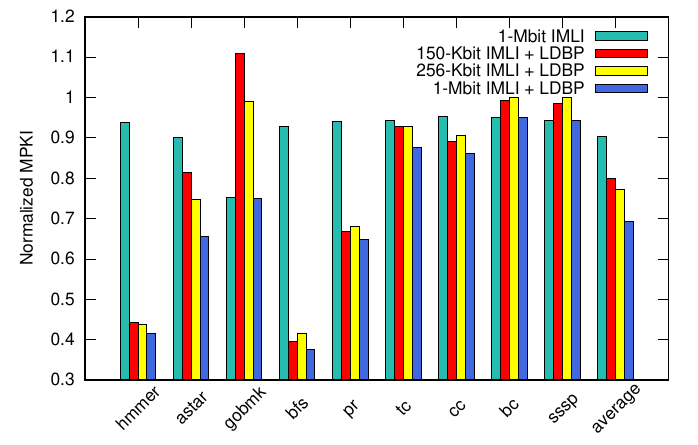}
%\caption[]{Normalized MPKI compared to the baseline 256-Kbit IMLI}
\caption[]{LDBP minimizes the mispredictions by more than 22.7\% when combined with the baseline 256-Kbit IMLI.}
\label{fig:mpki_percent_reduction}
\end{figure}

\section{Results and Analysis}
\label{sec:evaluation}

%\subsection{Experimental Results}

In this section, we highlight the results of our study. We compare the performance, and
 misprediction rate variations between the baseline IMLI predictor and our proposed LDBP
 predictor augmented to IMLI. Mispredictions Per Kilo Instruction (MPKI) is the metric used
 to compare the misprediction rate in this section.

\begin{figure}[htpb]
\centering
\includegraphics[width=1\linewidth]{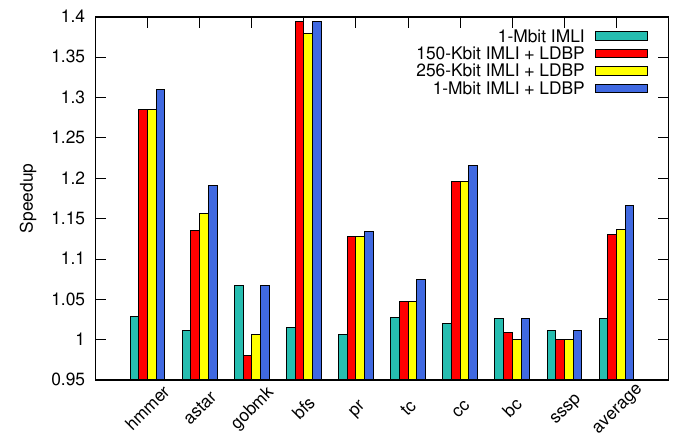}
%\caption[]{Normalized IPC over baseline 256-Kbit IMLI}
\caption[]{LDBP (when combined with 150-Kbit IMLI or 256-Kbit IMLI) outperforms the large 1-Mbit IMLI comprehensively.}
\label{fig:ipc_percent_increase}
\end{figure}

Figure~\ref{fig:mpki_percent_reduction} shows the normalized MPKI values compared to the baseline IMLI
for different branch predictor configurations. LDBP has a considerable impact on more than half
of the benchmarks. On average, the IMLI 256-Kbit + LDBP predictor reduces the MPKI of GAP and
SPEC CINT2006 benchmarks by 17.9\% and 27.5\%, respectively. As shown in
Table~\ref{tab:setup_table}, \textit{astar} had the worst branch prediction accuracy in the SPEC
CINT2006 suite. The most mispredicting branch in \emph{astar} constitutes 22\% of the benchmark's
mispredictions. This branch has a direct dependency with a load, but LDBP cannot fix this branch
as the address of the load feeding this branch has a fluctuating delta. LDBP manages to minimize \emph{astar's}
total branch misses by 25.4\% without fixing the most mispredicting branch. These numbers attest to the fact that
a considerable proportion of hard-to-predict branches on most benchmarks depend on data from loads
with a predictable address. Another observation to note is that quadrupling the size of IMLI fixes
only 9.7\% branch misses from the baseline. This inference substantiates the fact that a huge
TAGE-like predictor cannot efficiently capture the history of hard-to-predict data-dependent branches.

Figure~\ref{fig:ipc_percent_increase} compares IPC changes over baseline 256-Kbit IMLI for different
branch predictor configurations. LDBP was able to achieve an average IPC improvement of 13.7\% when paired
with the baseline predictor. An interesting observation is that the GAP benchmarks have a speedup of
12.5\% with this configuration. In contrast, they have a slightly better IPC gain of 12.9\% over the
baseline when running on 150-Kbit IMLI + LDBP. The reason for this trend is that a smaller IMLI can
fix lesser branches, and LDBP fixes branches that have low confidence with IMLI. Therefore, lower
the MPKI of the primary predictor, more the work for LDBP. A 41\% smaller IMLI (150-Kbit) with LDBP
produced similar IPC and MPKI numbers to that of the baseline IMLI-LDBP combination. For some benchmarks
like \emph{bfs}, the smaller predictor even outperformed its larger counterpart. Moreover, the 150-Kbit
IMLI + 81-Kbit LDBP offers 13.1\% higher performance gain and 20\% lesser branch misses than the baseline
256-Kbit IMLI for a 9.7\% lower hardware budget.

The MPKI and performance improvements yielded by LDBP clearly shows that
hard-to-predict load-dependent branches are major contributors to overall
mispredictions in benchmarks across different application suites. LDBP does not
affect some benchmarks like \emph{gobmk}, \emph{sssp} and \emph{bc}. This
behavior can be attributed that mispredicting branches in these benchmarks do
not have a load-branch dependency that can be captured by LDBP. An anomaly to
note on Figure~\ref{fig:mpki_percent_reduction}
and~\ref{fig:ipc_percent_increase} is the behavior of \emph{gobmk} running with
150-Kbit IMLI and LDBP. We can notice that the IPC decreases by 2\%, and the
MPKI worsens by 10\%. It is because the 150-Kbit IMLI has a worse MPKI and IPC
compared to the baseline 256-KBit IMLI. Added to that, LDBP does not yield any
improvement for \emph{gobmk}.

% PLOT: Main results. MPKI LDBP vs IMLI 256K vs IMLI 1M

% PLOT: Speedup

\begin{minipage}{0.45\textwidth}
\centering
\begin{lstlisting}[frame=single, caption={GAP BFS RISC-V Assembly for Listing~\ref{lst:bfs_cpp}}, label=lst:bfs]
L2: addi a7,a7,1
    ld	a5,8(t5)
    bge	a7,a5,L1 //outer 'for' loop
    sext.w t6,a7
    slli t1,a7,0x2
    ld	a5,0(a1)
    add	t1,t1,a5
    lw	a5,0(t1)
    bgez a5,L2  //'if' condition check
\end{lstlisting}
\end{minipage}

\subsection{Benchmark Study}

% code samples (asemply) for things that work in RISC-V

% For each bench pick 1 or 2 code snippets (something works and something that looks like work but it is not)
In this sub-section, we analyze examples from different benchmarks where LDBP works
and cases where LDBP doesn't work.

\subsubsection{Case 1: BFS (GAP Benchmark Suite)}

For our first case study, we look at the Breadth-First Search (BFS) algorithm. It is
one of the most popular graph traversal algorithms used across several
domains. Listing~\ref{lst:bfs} and Listing~\ref{lst:bfs_cpp} shows a snippet of
RISC-V assembly and its corresponding pseudo code from GAP's BFS benchmark. Here, the
loop traverses over all the nodes in the graph to assign a parent to each node. The
arbitrary nature of the graph makes it hard to predict if a node has a valid parent
as each node can have multiple possible edges, but the node traversal is in order. It
is hard to predict $parent[u]$, but $u$ is easily predictable (Line 2 in
Listing~\ref{lst:bfs_cpp}). The branch in Line 9 in Listing~\ref{lst:bfs} is the most
mispredicted branch in this benchmark. It contributes to about 30\% of all mispredictions
when simulated on the baseline architecture with 256-Kbit IMLI\@. When we augment LDBP
into this setup, it resolves about 94\% for the mispredictions of this branch and reduces
the overall MPKI by 59\%. It is also instrumental in gaining 38\% speedup.

\begin{minipage}{0.45\textwidth}
\centering
\begin{lstlisting}[language=c++, frame=single, caption={GAP BFS Source Code Snippet}, label=lst:bfs_cpp]
for(NodeID u=0; u < g.num_nodes(); u++){
  if(parent[u] < 0){
      ..
      ..
  }
}
\end{lstlisting}
\end{minipage}

\subsubsection{Case 2: HMMER (SPEC CINT 2006)}

Listing~\ref{lst:hmmer} shows the RISC-V assembly code section of the branch (line 8) contributing to
most misprediction in SPEC CINT2006 \emph{hmmer}. It accounts for 39\% of all mispredicted branches. The
branch outcome is dependent on values from different matrices. The randomness of the data involved makes
this a very hard-to-predict branch. Each branch source operand is dependent on two loads. As we traverse
over matrices, the loads involved in this case have a traceable address pattern. LDBP has to track four
different loads and some intermediate ALU operations to make the prediction. LDBP fixes 67\% of the
mispredictions yielded by \emph{bge}. Appending LDBP to the baseline IMLI improves the IPC by 29\% and
reduces the overall MPKI of this benchmark by 56\%.

\begin{minipage}{0.45\textwidth}
\centering
\begin{lstlisting}[frame=single, caption={SPEC CINT2006 hmmer RISC-V Assembly}, label=lst:hmmer]
lw    s11, 0(a3)
lw    a3, 4(a7)
addw  a3, s11, a3
sw    a3, 0(t3)
lw    s10, 0(s10)
lw    s11, 4(t1)
addw  s11, s10, s11
bge   a3, s11, LABEL
\end{lstlisting}
\end{minipage}

\subsubsection{Case 3: CC (GAP Benchmark Suite)}

Listing~\ref{lst:cc} represents the code-snippet containing the branch (line 5) with most mispredictions
in the CC benchmark. It constitutes a little more than one-third of all mispredictions in this
benchmark. LDBP cannot capture this load-branch chain. At first glance, it might look like the branch
instruction's source operands are dependent on two loads. On deeper introspection, we notice that the
source operand (load address) of the \emph{lw} instruction (recipient) on line 5 is determined by the
load data of the previous \emph{lw} (donor) on line 1. We refer to such a dependency as a load-load
chain. Figure~\ref{fig:chain} represents a load-load chain.

\begin{minipage}{0.45\textwidth}
\centering
\begin{lstlisting}[frame=single, caption={GAP CC RISC-V Assembly}, label=lst:cc]
lw    a6, 0(a4)
slli  a5, a6, 0x2
add   a5, a5, a0
lw    a5, 0(a5)
beq   a6, a5, LABEL
\end{lstlisting}
\end{minipage}

The current LDBP setup does not support load-branch slices having a load-load chain. If the address of the
first load instruction is predictable by the stride predictor, we can use its data to prefetch the second
load. Similar to the backward slice computation of the load-branch chain, we need to build a backward chain
starting from the recipient load and ending with the donor. If the load-load chain is predictable, then LDBP
can build the load-branch slice and generate predictions. This implementation is a part of our immediate future work.

 \begin{table}[htb]
   \resizebox{\columnwidth}{!}{%
     \begin{tabular} { c|cSS[table-format=2.2] }
     %\begin{tabular} { |c|c|c|c| }
    \textbf{Structure Name} & \textbf{No. of Entries} & \textbf{Total Size (Kbit)} \\
    \toprule
    Stride Predictor  &  48 & 2.39 \\
    Rename Tracking Table (RTT) & 32 & 3.09  \\
    Pending Load Queue (PLQ) & 48 & 0.33 \\
    Branch Trigger Table (BTT) & 8 & 0.88 \\
    Code Snippet Builder (CSB) & 32 & 4 \\
    Load Outcome Register (LOR) & 16 & 1.44 \\
    Load Outcome Table (LOT) & 16 &  65 \\
    Branch Outcome Table (BOT) & 8 & 1.93 \\
    Code Snippet Table (CST) & 8 & 2 \\
     \midrule
       \textbf{Total LDBP Size} &  & {\bf 81.06} \\
   \end{tabular}
   }
   \vspace{0.1in}
   \caption[]{Overall LDBP Size is 81-Kbit}
   \label{tab:ldbp_table_size}
 \end{table}

\subsection{LDBP Table Sizing}
\label{sec:table_sizing}
% Sizing structures

In this sub-section, we explain the methodology used to size the tables in LDBP. We analyze the
variation in MPKI for a different number of entries in each structure in the predictor. Here, MPKI
is the average MPKI of the benchmarks used. We define a baseline infinite LDBP predictor. The
infinite LDBP has 512 entries in each table. When the MPKI sensitivity for a table's size is
analyzed, all other tables in LDBP have 512 entries. Such an approach ensures a fair estimation
of the table's impact on LDBP accuracy. A 2\% MPKI increase from infinite LDBP is the cut-off
used to determine the ideal table size. Table~\ref{tab:ldbp_table_size} shows overall size of LDBP
and the breakdown of individual table sizes.

The overall size for the LDBP is 81-Kbit. As a reference, the IMLI predictor
used is 256-Kbit. The fetch block in a processor like a Zen 2 also includes a
32-KByte instruction cache and two-level BTBs with 512 and 7K entries. The largest
LDBP table is the LOT that can use area-efficient single port SRAMs.

\begin{figure}[htpb]
\centering
\includegraphics[width=0.8\linewidth]{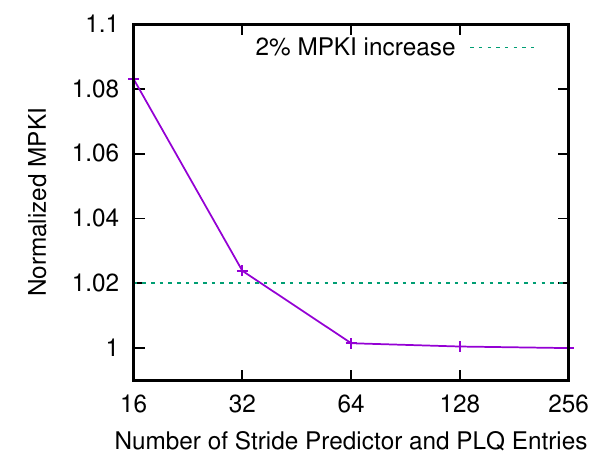}
%\caption[]{MPKI variation vs Stride Predictor/PLQ Entries}
\caption[]{48 entries are sufficient in the Stride Predictor and PLQ to achieve prediction accuracy varying by less than 2\% from the infinite LDBP.}
\label{fig:prefetch_sizing}
\end{figure}

 \subsubsection{Stride Predictor and PLQ Sizing}

Figure~\ref{fig:prefetch_sizing} shows the impact of the number of entries on the Stride Predictor
and the PLQ on MPKI. We can see that the MPKI drop is going over 2\% when the number of entries is
around 32. With reduced stride predictor and PLQ entries, a load tracked as a part of the
hard-to-predict load-dependent branch's chain can be evicted to make way for a new incoming
load. LDBP cannot determine if a load is trigger-worthy if it is not in the stride predictor
table. Entries larger than 64 have a negligible effect on the MPKI. The stride predictor and PLQ
have 48 entries each as it offers the perfect equilibrium between MPKI and hardware size.
%As shown in Table~\ref{tab:ldbp_table_size}, the stride predictor has a storage budget of 2.39-Kbits, and the PLQ occupies 0.33-Kbits.

%\begin{figure}[htpb]
%\centering
%\includegraphics[width=\linewidth]{plots/lor_sizing.pdf}
%\caption[]{MPKI variation vs LOR Entries}
%\caption[]{Tracking 16 loads on the LOR and LOT is adequate.}
%\label{fig:lor_sizing}
%\end{figure}

\begin{figure*}[ht]
  \centering
    \begin{minipage}[b]{0.3\textwidth}
      \centering
      \includegraphics[width=1\textwidth]{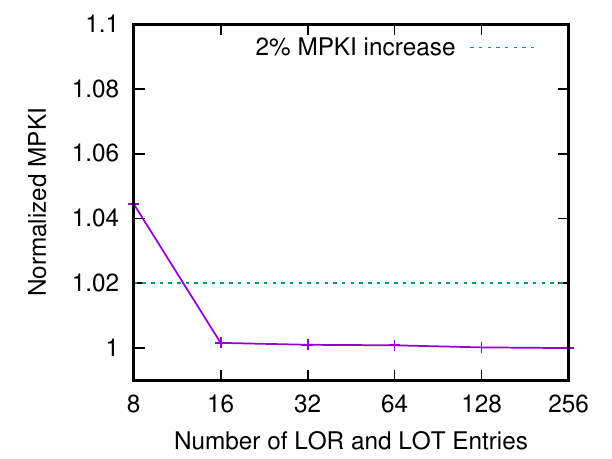}
      %\caption{MPKI vs Number of loads in LDBP chain}
      \caption{Tracking 16 loads on the LOR and LOT is adequate to maintain high LDBP accuracy.}
      \label{fig:lor_sizing}
    \end{minipage}
    \hspace{0.3cm}
    \begin{minipage}[b]{0.3\textwidth}
      \centering
      \includegraphics[width=1\textwidth]{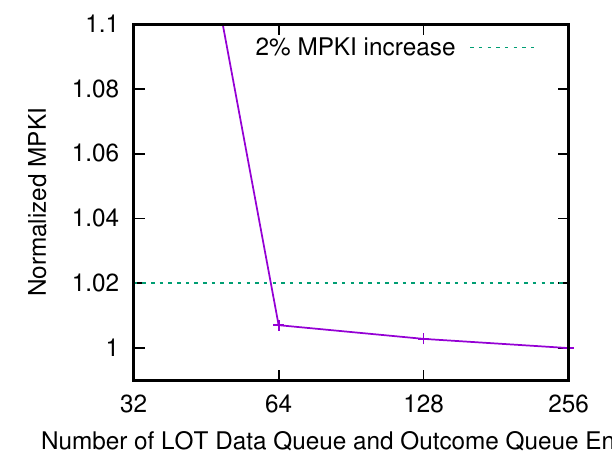}
      %\caption{MPKI vs Number of ALU ops in LDBP chain}
      \caption{The LOT Data Queue and Outcome Queue requires 64 entries each.}
      \label{fig:lotq_sizing}
    \end{minipage}
    \hspace{0.3cm}
    \begin{minipage}[b]{0.3\textwidth}
      \centering
      \includegraphics[width=1\textwidth]{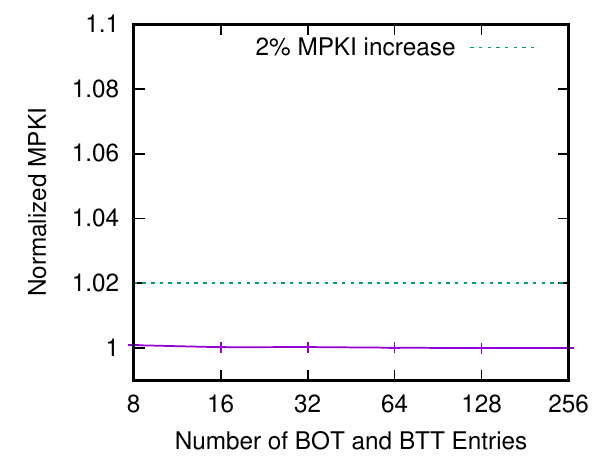}
      %\caption{MPKI vs Number of sub-entries in each CSB index}
      \caption{Effectiveness of LDBP remains steady for different number of entries on the BOT and BTT.}
      \label{fig:bot_sizing}
    \end{minipage}
\end{figure*}

\begin{figure*}[ht]
  \centering
    \begin{minipage}[b]{0.3\textwidth}
      \centering
      \includegraphics[width=1\textwidth]{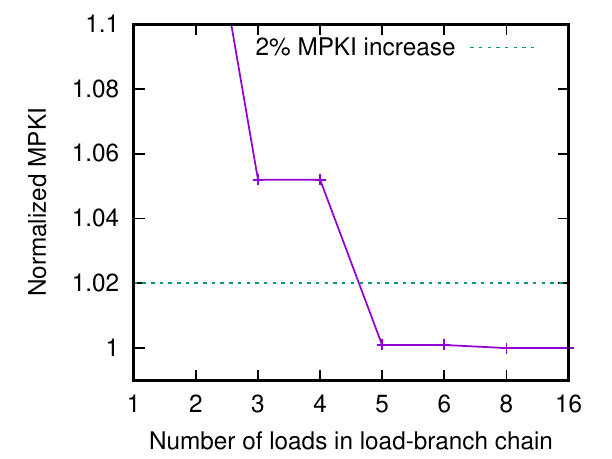}
      %\caption{MPKI vs Number of loads in LDBP chain}
      \caption{LDBP must track at least 5 loads to maintain healthy prediction accuracy.}
      \label{fig:num_lds}
    \end{minipage}
    \hspace{0.3cm}
    \begin{minipage}[b]{0.3\textwidth}
      \centering
      \includegraphics[width=1\textwidth]{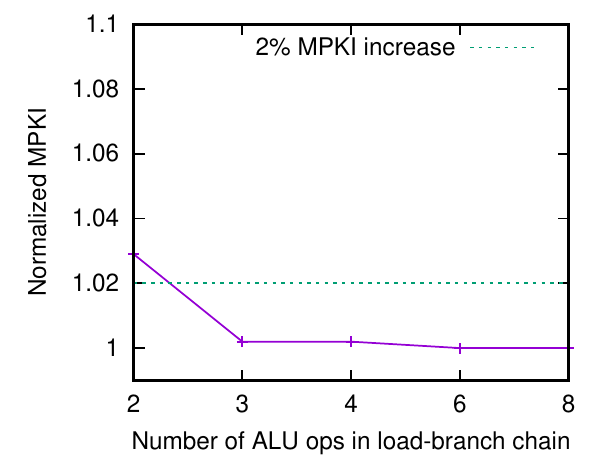}
      %\caption{MPKI vs Number of ALU ops in LDBP chain}
      \caption{Most LDBP chains have 3 ALU operations between the loads and branch.}
      \label{fig:num_ops}
    \end{minipage}
    \hspace{0.3cm}
    \begin{minipage}[b]{0.3\textwidth}
      \centering
      \includegraphics[width=1\textwidth]{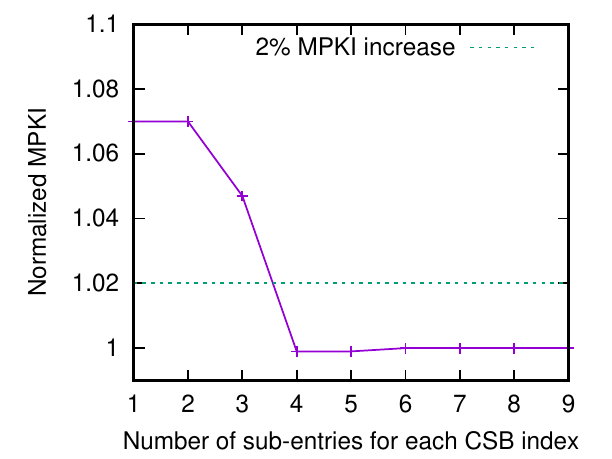}
      %\caption{MPKI vs Number of sub-entries in each CSB index}
      \caption{Each CSB index must have 4 sub-entries to capture LDBP backward slice.}
      \label{fig:csb_sizing}
    \end{minipage}
\end{figure*}

\subsubsection{LOR and LOT Sizing}

Figure~\ref{fig:lor_sizing} plots the effect of varying LOR size on MPKI. There
is a sharp increase in MPKI when the number of trigger loads tracked is less than 16. At the
2\% cut-off point, LOR and LOT has around 12 entries. To minimize the impact of the sharp drop in
MPKI, we allocate 16 entries to both the LOR and LOT. The necessity to store the complete load
data contributes to the large size of the LOT. The number of entries on the LOT data queue is
determined by how proactively LDBP wants to predict branches and trigger its associated loads. The
number of entries on the BOT's outcome queue matches the LOT data queue entries. The sizing of the
BOT outcome queue is discussed in Section~\ref{sec:oq_sizing}.

%\begin{figure}[htpb]
%\centering
%\includegraphics[width=\linewidth]{plots/lotq_sizing.pdf}
%\caption[]{MPKI variation vs Outcome Queue/LOT Data Queue Entries}
%\caption[]{The LOT Data Queue and Outcome Queue requires 64 entries each.}
%\label{fig:lotq_sizing}
%\end{figure}

Some load-dependent branches may consume two or more trigger loads. A bottleneck on the
number of trigger loads tracked has a direct implication on the effectiveness of LDBP. In
most cases, the load-dependent branch tends to be the entry-point to a huge loop. In such
cases, it sufficient for LDBP to track just one branch and its associated trigger
loads. Therefore, a reasonably small to medium number of entries on the LOR and LOT is adequate
to maintain LDBP accuracy.

%\begin{figure}[htpb]
%\centering
%\includegraphics[width=\linewidth]{plots/bot_sizing.pdf}
%\caption[]{MPKI variation vs BOT/BTT Entries}
%\caption[]{Prediction accuracy of LDBP remains steady across different number of entries on the BOT and BTT.}
%\label{fig:bot_sizing}
%\end{figure}

\subsubsection{Outcome Queue/LOT Data Queue Sizing}
\label{sec:oq_sizing}

The outcome queue is part of the BOT. The criticality of the outcome queue in the overall
scheme of LDBP warranted optimal sizing. The number of entries in this queue correlates to
the number of future outcomes trackable for a given branch PC. The outcome queue entries
directly impact the number of entries on the LOT data queue. It is sufficient for the LOT
data queue to have as many entries as the branches tracked by the outcome queue. From
Figure~\ref{fig:lotq_sizing}, the ideal number of outcome queue entries at the cut-off point
is 64. As the outcome queue size decreases, the MPKI increase gets steeper. A smaller outcome
queue inhibits the ability of LDBP to trigger loads with higher prefetch distance. On the flip
side, the outcome queue size larger than 64 almost hits an MPKI plateau.

\subsubsection{BOT and BTT Sizing}

Figure~\ref{fig:bot_sizing} shows the variation of MPKI for different sizes of BOT and BTT. Just
like the LOR and LOT, a small to a medium number of entries on the BOT and BTT is sufficient to
track almost every load-dependent branch in an application. These branches are usually part of
large loops. These huge loops give LDBP adequate time to capture the new branch-load chain even
if they replace an already existing entry from the tables. The correlation between the number of
entries and MPKI has very minimal variations. Therefore, it is sufficient to have just 8 entries
on the BOT and BTT.

\subsubsection{CSB and CST Sizing}
The CSB builds the load-branch slice. It is critical to size this table optimally to keep LDBP's
hardware budget under check. Figure~\ref{fig:num_lds} and~\ref{fig:num_ops} shows the change of
MPKI for different load and ALU operations threshold in an LDBP chain. Five loads and
three ALU operations are needed to ensure maximum LDBP efficiency. These figures reflect the
cumulative number of operations tracked by both the source operands of a branch instruction. Each
source operand of the branch might need to track only fewer operations.

Figure~\ref{fig:csb_sizing} shows the number of sub-entries needed by each CSB index. This figure
clearly shows that it is sufficient for each branch source register to track four operations to
support an LDBP chain with a maximum of eight operations. There are 32 entries on the CSB, and each
entry track four operations. The total size of the CSB is 4-Kbit. The CST caches the backward slice
of each branch. As there are 8 entries on the BOT, the CST must have 8 entries with 8 sub-entries
(4 sub-entry for each branch source operand).

%\subsection{Branch Predictor Size Implications}

%Table~\ref{tab:ldbp_table_size} consolidates the number of table entry values from
%Section~\ref{sec:table_sizing} to compute the overall size of the proposed LDBP
%predictor. Figure~\ref{fig:mpki_percent_reduction} and~\ref{fig:ipc_percent_increase}
%shows the MPKI and IPC behavior of different sized standalone IMLI
%predictors, respectively. Compared to the baseline 256-Kbit IMLI, the 1-Mbit IMLI offers
%only 2.7\% average speedup and an 9.7\% reduction in mispredictions across the investigated
%benchmarks. These numbers are not an ideal tradeoff for a 4x increase in the predictor
%budget. A 41\% smaller IMLI (150-Kbit) with LDBP produced similar IPC and MPKI numbers to
%that of the baseline IMLI-LDBP combination. For some benchmarks like \emph{bfs}, the smaller
%predictor even outperformed its larger counterpart. This observation clearly shows that the
%budget of the primary predictor does not perturb the potency of LDBP. Moreover, the 150-Kbit
%IMLI + 81-Kbit LDBP offers 13.1\% higher performance gain and 20\% lesser branch misses
%than the baseline 256-Kbit IMLI for a 9.7\% lesser hardware budget.

\subsection{LDBP Gating and Energy Implications}

The LDBP has significant performance gains, but some benchmarks (\emph{gobmk}, \emph{sssp},
\emph{bc}) do not benefit. We evaluate the effectiveness of gating the LDBP when
infrequently used, to save energy consumption.

We gate (low-power mode) every component of LDBP apart from the Stride
Predictor and RTT when there is a duration of 100,000 or more clock cycles
where LDBP did not predict any branch. We refer to this phase as the LDBP
low-power mode. As shown in Table~\ref{tab:ldbp_gating}, for \emph{bc} and
\emph{sssp}, LDBP remains in low-power mode for 99.5\% and 98.2\% of the
benchmark's execution time, respectively. Gating offers a considerable
reduction in energy dissipated by LDBP as the predictor remains in low-power
mode for 38.5\% of the average execution time across all benchmarks, and LDBP
gating does not have any negative effect on the prediction accuracy of LDBP.

 \begin{table}[htb]
   \centering
   \small
   %\resizebox{\columnwidth}{!}{%
     \begin{tabular} { c|S[table-format=2.1] }
       \textbf{Benchmark} & \textbf{\% time in low-power mode} \\
    \toprule
    spec06\_hmmer  &  0.0 \\
    spec06\_astar  &  54.6 \\
    spec06\_gobmk  &  36.1 \\
     \midrule
    gap\_bfs  &  4.0 \\
    gap\_pr  &  2.8 \\
    gap\_tc  &  0.2 \\
    gap\_cc  &  50.8 \\
    gap\_bc  &  99.5 \\
    gap\_sssp  &  98.2 \\
   \end{tabular}
   %}
   \vspace{0.1in}
   \caption[]{Proportion of execution time in LDBP low-power mode}
   \label{tab:ldbp_gating}
 \end{table}

Haj-Yihia et al.~\cite{haj2016fine} present a detailed breakdown of core power consumption for high-performance modern
CPUs running SPEC CINT2006 benchmarks. We use the data presented in their work to estimate the core energy dissipation
for our analysis. For our baseline energy model, we replicate the core power breakdown given in \cite{haj2016fine} for
\emph{hmmer}, \emph{astar} and \emph{gobmk}. For the GAP benchmarks, we use the average power breakdown of SPEC CINT2006
benchmarks given in \cite{haj2016fine}. The broad-spectrum power model based on SPEC CINT2006 benchmarks is good enough
to capture the energy dissipation behavior of GAP benchmarks with a good level of accuracy.

\begin{figure}[htpb]
\centering
\includegraphics[width=0.83\linewidth]{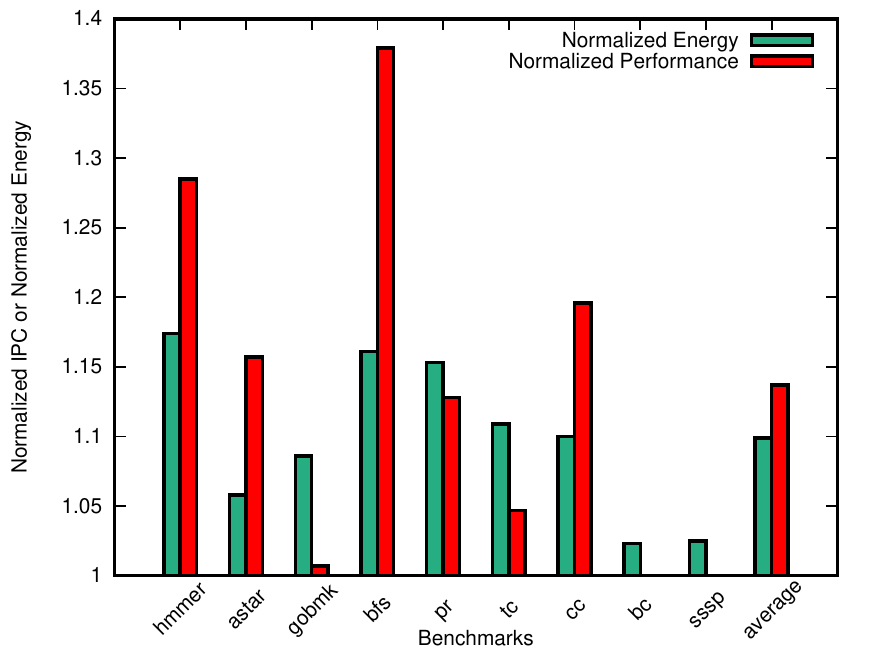}
%\caption[]{IPC variation of IMLI and LDBP for different core optimizations}
\caption[]{LDBP maintains a favorable energy-performance tradeoff.}
\label{fig:energy_ipc}
\end{figure}

Energy Per Access (EPA) for IMLI and LDBP were calculated using CACTI
6.0~\cite{muralimanohar2009cacti}. For IMLI, we model an ideal structure with a
single port. LDBP has 55\% lesser EPA than IMLI even if we assume all the
tables are accessed when not in low power mode, which is not the case in
reality. There is a 10.9\% average increase in DL1 access for LDBP, which will
result in an equivalent escalation in energy on the memory sub-system. The
22.7\% decrease in MPKI when using LDBP will compensate for this increase in
energy dissipation. Lesser MPKI implies lesser energy spent on executing the
wrong branch path. We do not account for the energy saved due to reduced wrong
path execution in the LDBP energy estimation numbers. Added to that, we also do
not account for the energy reduction incurred due to 13.3\% lesser execution
time when using LDBP. Reducing execution time results in reduced energy, and
our pessimistic energy estimation model for LDBP does not consider this.

Figure~\ref{fig:energy_ipc} shows the energy-performance tradeoff for IMLI +
LDBP compared to the baseline 256-Kbit IMLI. The IPC boost outweighs the
increase in energy dissipation for the majority of the benchmarks that benefit
from LDBP. Benchmarks like \emph{bc} and \emph{sssp} only have about 2\% energy
overhead as the RTT and Stride Predictor continue to be active even under
low-power mode. Interestingly, LDBP only predicts a negligible proportion of
branches in \emph{gobmk}, but it contributes to 8\% more energy use. This is
because LDBP resolves multiple low-frequency branches that spread across
different execution phases. Thus, \emph{gobmk} does not offer a consistent
low-power mode phase for LDBP. A more aggressive clock gating with retention
state or smarter phase learning could further improve the \emph{gobmk} case, but we
leave it as future work.

%\subsection{Impact of Prefetching and Memory Bottleneck on LDBP}
\subsection{Impact of Triggering Loads on LDBP Performance Gains}

\begin{figure}[htpb]
\centering
\includegraphics[width=0.83\linewidth]{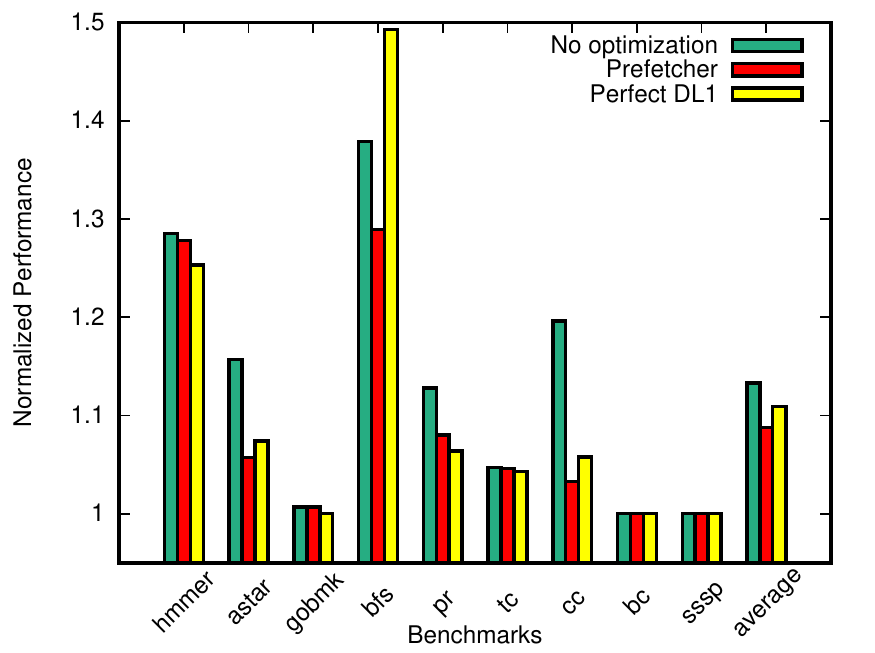}
%\caption[]{IPC variation of IMLI and LDBP for different core optimizations}
\caption[]{Triggering loads does not offer any unfair gains to LDBP.}
\label{fig:ldbp_prefetch_impact}
\end{figure}

Figure~\ref{fig:ldbp_prefetch_impact} shows the normalized speedup of LDBP over 256-Kbit IMLI with
three different Zen 2 core configurations. One, the default Zen-2 core used for
evaluation in other parts of this paper. Two, the default core with a standard
stride prefetcher and third, the default core with perfect DL1 cache. We can
notice that the IPC numbers are almost similar across all three configurations
for most benchmarks. This clearly shows that prefetching trigger loads in LDBP
do not provide an unfair advantage to it over the standalone IMLI predictor.
Maybe even more important, Figure~\ref{fig:ldbp_prefetch_impact} shows that the LDBP benefits are
consistent independent of memory sub-system improvements.

%The IPC of \emph{astar} and \emph{cc} are constrained by a high misprediction rate as well as memory
%For \emph{astar} and \emph{cc}, the standalone IMLI with prefetching or perfect DL1 outperforms its unoptimized counterpart
%by easing the bottleneck on memory. Therefore, the IMLI+LDBP with prefetching or perfect DL1 has a
%relatively lower speedup compared to its unoptimized equivalent. In \emph{bfs}, high MPKI and memory
%bottleneck restrict IPC. LDBP has very high effectiveness with BFS as the top mispredicted branches are
%dependent on loads. Therefore, a core with perfect DL1 and LDBP has 10\% more speedup than LDBP with no optimization.

%\begin{minipage}{0.45\textwidth}
%\centering
%\begin{lstlisting}[frame=single, caption={Pseudo-code of Listing~\ref{lst:asm} (vector traversal example)}, label=lst:trivial]
%int total = 0;
%for(int v = 0; v < vec.size(); v++)
%  if (vec[v] < 0)
%    total = total + 1;
%\end{lstlisting}
%\end{minipage}

% Cacti table power area vs IMLI base
\subsection{Trigger Load Timeliness}

In this sub-section, we will focus on trigger load prefetch distance and its importance in achieving
optimum LDBP timeliness. We will use Listing~\ref{lst:asm} to highlight the criticality of timely trigger
loads. This example is the vector traversal problem discussed in Section~\ref{sec:intro}. In the
example we discuss, let us assume a scenario where it takes six cycles to load data from the vector, and
there are ten inflight load-branch iterations. As the load address has a delta of 8, to achieve an IPC of
1, we need to send the new trigger load at least 16 cycles ahead. If the current load address is $x$, LDBP
triggers a load address with a distance of 16 ($x + 8 * 16$). In reality, it would be ideal to use even a
larger distance to compensate for variable memory latencies. Larger trigger distances require more buffering
and can be potentially more wasteful if the stride pattern changes. Triggering too far ahead can also pollute
the cache and could evict useful lines. Another vital point to note is, a trigger load is an actual load and
not a prefetch. We overlooked the idea of using prefetches in LDBP as it has the potential to be dropped.

\begin{figure}[htpb]
\centering
\includegraphics[width=0.83\linewidth]{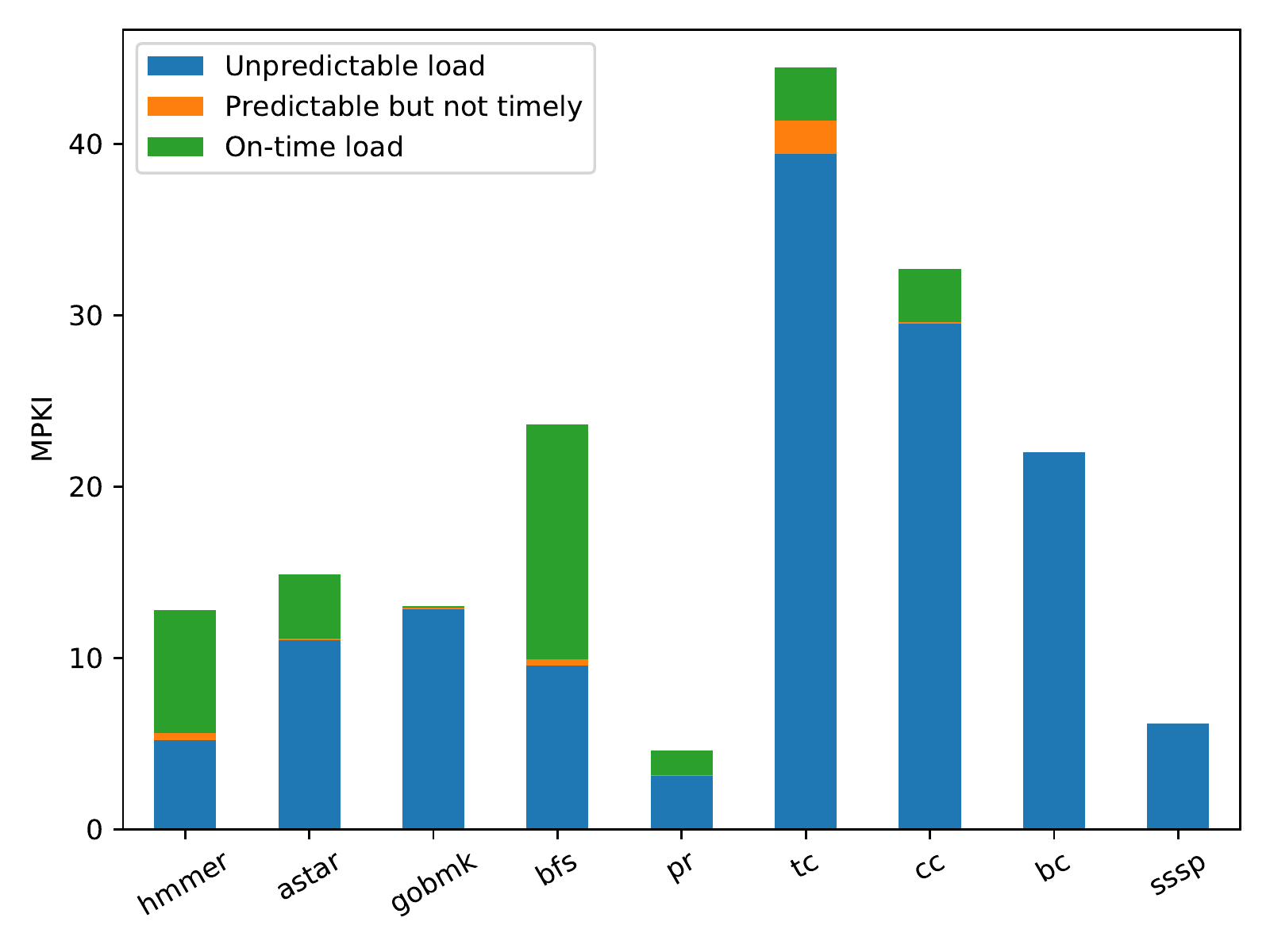}
\caption[]{MPKI variation vs Trigger Load Timeliness.}
\label{fig:load_timeliness}
\end{figure}

Figure~\ref{fig:load_timeliness} shows the MPKI of different benchmarks
recorded with the baseline IMLI predictor. The portion of the bar shaded in
green points to the number of mispredictions fixed by LDBP due to the timely
execution of trigger loads. We can see the correlation between the number of
predictable loads in a benchmark and LDBP effectiveness. The timeliness of
these predictable load helps to exploit the maximum potency of LDBP. Only for
the \emph{tc} benchmark, a significant portion of the loads are not predictable.
Though we optimized the methodology to trigger loads, these outliers can be
attributed to the change in load delta, which creates considerable delay due to
relearning time. Another potential reason could be memory bandwidth congestion.
Minimizing the number of delayed trigger loads could lead to significant MPKI
reduction.

% PLOT: Slice delay

% PLOT: Breakdown MPKI (timely, predictable but late, not-predictable)

\section{Related Work}
\label{sec:related}

The strides in branch prediction accuracy have improved several folds since the counter-based bimodal
predictor~\cite{smith81bps}. The ensuing works on branch prediction gradually raised the bar for the
prediction accuracy. Yeh and Patt came up with the two-level branch predictors~\cite{yeh91two}. McFarling~\cite{mcfarling93}
proposed optimizations over their work. These works leverage the high correlation between the outcome
of the current branch and the history of previous branch outcomes.

PPM-like~\cite{michaud2005ppm} and TAGE~\cite{seznec2006case} achieve higher prediction accuracy by
tracking longer histories. They use multiple prediction tables, each indexed by a longer global history
than its preceding table. TAGE-based predictors are the state-of-the-art predictors, and they offer very
high prediction accuracy. TAGE-based predictors fail to capture the outcome correlation of branches having
an irregular periodicity or when a branch outcome history is too long or too random to capture.

Statistical correlator~\cite{seznec201164} and IMLI~\cite{seznec2015inner} components are augmented to TAGE
to mitigate some of the mispredictions. Several studies and extensive workload analysis have identified different
types of hard-to-predict branches and ways to resolve them. Sherwood et al.~\cite{sherwood2000loop} and
Morris et al.~\cite{morris2002loop} proposed prediction mechanisms to tackle loop-termination branches. The Wormhole
predictor~\cite{albericio2014wormhole} improved on earlier loop-based predictors to handle branches enclosed within
nested loop and branches exhibiting correlation across different iterations of the outer loop.

Branches dependent on random data from load instructions contribute to a high percentage of mispredictions with
TAGE-bases predictors. It is impossible to capture the history of such branches competently, even with an unusually
large predictor. Prior works~\cite{heil1999impr, chen2003dynamic} show that using data values as an input to the
branch predictor improves the misprediction rate. Farooq et al.~\cite{farooq2013store} note that some hard-to-predict
data-dependent branches manifest a specific pattern of a store-load-branch chain. They leverage this observation to
mark the stores that are in the chain at compile-time and compute branch conditions based on the values of marked
stores at run-time in hardware. We tackle a similar problem, but our work is based on the observation that a
considerable proportion of hard-to-predict data-dependent branches are dependent on the loads whose address is very
predictable. Moreover, we do not make any modifications to the ISA. Gao et al.~\cite{gao2008address} proposed a
closely related work. They correlate the branch outcome to the load address and provide a prediction based on the
confidence of the correlation. Nevertheless, our approach differs in that we precalculate the branch outcomes by
triggering loads that are part of the branch's dependence chain and have a highly predictable address.

%\section{Conclusions and Future Work}
\section{Conclusions}
\label{sec:conclusions}

As shown by the benchmarks evaluated in our work, branch outcomes dependent on arbitrary
load data are hard-to-predict and contribute to most mispredictions. They have poor
prediction accuracy with current state-of-the-art branch predictors. These branch patterns
are common in data structures like vector, maps, and graphs. We propose the Load Driven
Branch Predictor (LDBP) to eliminates the misses contributed by this class of branch. LDBP
exploits the predictable nature of the address of the loads on which these hard-to-predict
branches depend on and triggers these dependent loads ahead of time. The triggered load data
are used to precompute the branch outcome. With LDBP, programmers can traverse over large
vectors/maps, do data-dependent branches, and still have near-perfect branch prediction.

LDBP contributes to minimal hardware and power overhead and does not require any changes to
the ISA. Our experimental results show that compared to the standalone 256-Kbit IMLI
predictor, the combination of 256-Kbit IMLI and LDBP predictor shrinks the branch MPKI by 22.7\%
and improves the IPC by 13.7\%. The efficiency of LDBP also allows having a smaller primary
predictor. A 150-Kbit IMLI + LDBP predictor yields performance improvement of 13.1\% and 20\%
lesser mispredictions compared to the baseline 256-Kbit IMLI.

%LDBP is a new class of branch prediction that enables not only improvements in
%current software, but allows changes to existing software to further enhance
%performance. With LDBP, programmers can traverse over large vectors/maps, do data
%dependent branches, and still have near-perfect branch prediction.

Another opportunity that this work provides is to extend the use of graphs
further. As the GAP benchmark suite results show, LDBP can improve performance
from graph traversals significantly. There is an extensive set of works exploring
graphs for neural networks~\cite{wu2019comprehensive}, for which LDBP could help
to boost the performance.

%Another opportunity that this work provides is to further extend the use of
%graphs. As the GAP benchmark suite results show, LDBP can improve performance from graph
%traversals significantly. There is an extensive set of work exploring graph for
%neural networks~\cite{wu2019comprehensive}, LDBP could help the performance of
%sparse graph algorithms.

%\section{Conclusions and Future Work}
\section{Acknowledgements}
\label{sec:acknowledgements}

This work was supported in part by the National Science Foundation under grant CCF-1514284.
Any opinions, findings, and conclusions or recommendations expressed herein are
those of the authors and do not necessarily reflect the views of the NSF.
This material is based upon work supported by, or in part by, the Army Research
Laboratory and the Army Research Office under contract/grant W911NF1910466.

%%%%%%% -- PAPER CONTENT ENDS -- %%%%%%%%

%%%%%%%%% -- BIB STYLE AND FILE -- %%%%%%%%
\bibliographystyle{IEEEtran}
%\bibliography{refs}
\bibliography{ms.bib}
%%%%%%%%%%%%%%%%%%%%%%%%%%%%%%%%%%%%

\end{document}